%% file: main.tex
\documentclass[sigconf]{acmart}
\pdfoutput=1
\renewcommand\footnotetextcopyrightpermission[1]{}
\acmConference{}{}{ }

\usepackage{tugraz_defaults}
\usepackage{csquotes}
\usepackage{enumitem}
\usepackage{longtable}

\renewcommand{\subparagraph}[1]{\vspace{0.0cm}\noindent\textbf{#1}\ }

\newcommand{\Speechminer}{\textsc{SpeechMiner}\xspace}

\usepgfplotslibrary{dateplot}
\usepgfplotslibrary{fillbetween}

\newcommand{\RegAttack}{\emph{Dereference Trap}\xspace}

\begin{document}
\title{Speculative Dereferencing of Registers: Reviving Foreshadow}

\author{Martin Schwarzl}
\affiliation{Graz University of Technology}
\email{martin.schwarzl@iaik.tugraz.at}

\author{Thomas Schuster}
\affiliation{Graz University of Technology}
\email{thomas.schuster@student.tugraz.at}

\author{Michael Schwarz}
\affiliation{CISPA Helmholtz Center for Information Security}
\email{michael.schwarz@cispa.saarland}

\author{Daniel Gruss}
\affiliation{Graz University of Technology}
\email{daniel.gruss@iaik.tugraz.at}



\begin{abstract}
Since 2016, multiple microarchitectural attacks have exploited an effect that is attributed to prefetching.
These works observe that certain user-space operations can fetch kernel addresses into the cache.
Fetching user-inaccessible data into the cache enables KASLR breaks and assists various Meltdown-type attacks, especially Foreshadow.

In this paper, we provide a systematic analysis of the root cause of this prefetching effect.
While we confirm the empirical results of previous papers, we show that the attribution to a prefetching mechanism is fundamentally incorrect in all previous papers describing or exploiting this effect.
In particular, neither the \texttt{prefetch} instruction nor other user-space instructions actually prefetch kernel addresses into the cache,\footnote{Various authors of papers exploiting the prefetching effect confirmed that the explanation put forward in this paper indeed explains the observed phenomena more accurately than their original explanations. We believe it is in the nature of empirical science that theories explaining empirical observations improve over time and root-cause attributions become more accurate.} leading to incorrect conclusions and ineffectiveness of proposed defenses.
The effect exploited in all of these papers is, in fact, caused by speculative dereferencing of user-space registers in the kernel.
Hence, mitigation techniques such as KAISER do not eliminate this leakage as previously believed.
Beyond our thorough analysis of these previous works, we also demonstrate new attacks enabled by understanding the root cause, namely an address-translation attack in more restricted contexts, direct leakage of register values in certain scenarios, and the first end-to-end Foreshadow (L1TF) exploit targeting non-L1 data.
The latter is effective even with the recommended Foreshadow mitigations enabled and thus revives the Foreshadow attack.
We demonstrate that these dereferencing effects exist even on the most recent Intel CPUs with the latest hardware mitigations, and on CPUs previously believed to be unaffected, \ie ARM, IBM, and AMD CPUs.
\end{abstract}

\maketitle
\pagestyle{plain}



\section{Introduction}
Modern system security depends on isolating domains from each other.
One domain cannot access information from the other domain, \eg another process or the kernel.
Hence, the goal of many attacks is to break this isolation and obtain information from other domains.
Microarchitectural attacks like Foreshadow~\cite{Vanbulck2018foreshadow,Weisse2018foreshadow} and Meltdown~\cite{Lipp2018meltdown} gained broad attention due to their impact and mitigation cost.
One building block that facilitates microarchitectural attacks is knowledge of physical addresses.
Knowledge of physical addresses can be used for various side-channel attacks~\cite{Liu2015Last,Irazoqui2015SA,Maurice2015C5,Gruss2016Flush,Pessl2016}, bypassing SMAP and SMEP~\cite{Kemerlis2014}, and mounting Rowhammer attacks~\cite{Seaborn2015BH,Kim2014,Xiao2016,Bhattacharya2016,Razavi2016,Islam2019Spoiler}.
As a mitigation to these attacks, operating systems do not make physical address information available to user programs~\cite{linux_patch_pagemap}.
Hence, the attacker has to leak the privileged physical address information first.
The \emph{address-translation attack} by Gruss~\etal\cite{Gruss2016Prefetch} solves this problem.\footnote{This attack is detailed in Section 3.3 and Section 5 of the Prefetch Side-Channel Attacks paper~\cite{Gruss2016Prefetch}}
The address-translation attack allows unprivileged applications to fetch arbitrary kernel addresses into the cache and thus resolve virtual to physical addresses on 64-bit Linux systems.
As a countermeasure against microarchitectural side-channel attacks on kernel isolation, \eg the address-translation attack, Gruss~\etal\cite{Gruss2016Prefetch,Gruss2017KASLR} proposed the KAISER technique.

More recently, other attacks observed and exploited similar prefetching effects.
Lipp~\etal\cite{Lipp2018meltdown} described that Meltdown successfully leaks memory that is not in the L1 cache, but did not thoroughly explain why this is the case.
Xiao~\etal\cite{Xiao2019Speechminer} show that this is only possible due to a prefetching effect, when performing Meltdown-US, where data is fetched from the L3 cache into the L1 cache.
Van Bulck~\etal\cite{Vanbulck2018foreshadow} observe that for Foreshadow this effect does not exist.
Foreshadow is still limited to the L1, however in combination with Spectre gadgets which fetch data from other cache levels it is possible to bypass current L1TF mitigations.
This statement was further mentioned as a restriction by Canella~\etal\cite{Canella2019A} and Nilsson~\etal\cite{Nilsson2020SGXSurvey}.
Van Schaik~\etal state that Meltdown is not fully mitigated by L1D flushing~\cite{VanSchaik2019RIDL}.

We systematically analyze the root cause of the prefetching effect exploited in these works.
We first empirically confirm the results from these papers, underlining that these works are scientifically sound, and the evaluation is rigorous.
We then show that, despite the scientifically sound approach of these papers, the attribution of the root cause, \ie why the kernel addresses are cached, is incorrect in all cases.
We discovered that this prefetching effect is actually unrelated to software prefetch instructions or hardware prefetching effects due to memory accesses and instead is caused by speculative dereferencing of user-space registers in the kernel.
While there are multiple code paths which trigger speculative execution in the kernel, we focus on a code path containing a Spectre-BTB~\cite{Kocher2019,Canella2019A} gadget which can be reliably triggered on both Linux and Windows.

Based on our new insights, we correct several assumptions from previous works and present several new attacks exploiting the underlying root cause.
We demonstrate that an attacker can, in certain cases, observe caching of the address (or value) stored in a register of a different context.
Based on this behavior, we present a cross-core covert channel that does not rely on shared memory.
While Spectre ``prefetch'' gadgets, which fetch data from the last-level cache into higher levels, are known~\cite{Canella2019A}, we show for the first time that they can directly leak actual data.
Schwarz~\etal\cite{Schwarz2019ZL} showed that prefetch gadgets can be used as a building block for ZombieLoad on affected CPUs to not only leak data from internal buffers but to leak arbitrary data from memory.
We show that prefetch gadgets are even more powerful by also leaking data on CPUs unaffected by ZombieLoad.
Therefore, we demonstrate for the first time data leakage with prefetch gadgets on non-Intel CPUs.

The implications of our insights affect the conclusions of several previous works.
Most significantly, the difference that Meltdown can leak from L3 or main memory~\cite{Lipp2018meltdown} but Foreshadow (L1TF) can only leak from L1~\cite{Vanbulck2018foreshadow}\footnote{Appendix Foreshadow's Cache Requirement~\cite{Vanbulck2018foreshadow} and subsequently also reported by Canella~\etal\cite{Canella2019A} (Table 4~\cite{Canella2019A}), and Nilsson~\cite{Nilsson2020SGXSurvey} (Section III.E~\cite{Nilsson2020SGXSurvey}.}, was never true in pratice.
For both, Meltdown and Foreshadow, the data has to be fetched in the L1 to get leaked.
However, this restriction can be bypassed by exploiting prefetch gadgets to fetch data into L1.
Therefore L1TF was in practice never restricted to the L1 cache, due to the same ``prefetch'' gadgets in the kernel and hypervisor that were exploited in Meltdown.
Because of these gadgets, mounting the attack merely requires moving addresses from the hypervisor's address space into the registers.
Hence, we show that specific results from previous works are only reproducible on kernels that still have such a ``prefetch'' gadget, including, \eg Gruss~\etal\cite{Gruss2016Prefetch},\footnote{The address-translation oracle in Section 3.3 and Section 5 of the Prefetch Side-Channel Attacks paper~\cite{Gruss2016Prefetch}.} Lipp~\etal\cite{Lipp2018meltdown},\footnote{The L3-cached and uncached Meltdown experiments in Section 6.2~\cite{Lipp2018meltdown}.}, Xiao~\etal\cite{Xiao2019Speechminer}\footnote{The L3-cached experiment in Section IV-E~\cite{Xiao2019Speechminer}.}.
We also show that van Schaik~\etal\cite{VanSchaik2019RIDL} (Table III~\cite{VanSchaik2019RIDL}) erroneously state that L1D flushing does not mitigate Meltdown.

We then show that certain attacks can be mounted in JavaScript in a browser, as the previous assumptions about the root cause were incorrect.
For instance, we recover physical addresses of a JavaScript variable to be determined with cache-line granularity~\etal\cite{Gruss2016Prefetch}.
Knowledge of physical addresses of variables aids Javascript-based transient-execution attacks~\cite{Kocher2019,Mcilroy2019}, Rowhammer attacks~\cite{Gruss2016Row,Islam2019Spoiler}, cache attacks~\cite{Oren2015}, and DRAMA attacks~\cite{Schwarz2017Timers}. 

We then show that we can mount Foreshadow attacks on data not residing in L1 on kernel versions containing ``prefetch'' gadgets.
Worse still, we show that for the same reason Foreshadow mitigations~\cite{Vanbulck2018foreshadow,Weisse2018foreshadow} are incomplete.
We reveal that a full mitigation of Foreshadow attacks additionally requires Spectre-BTB mitigations (\texttt{nospectre\_v2}), a fact that was not known or documented so far.

We demonstrate that the prefetch address-translation attack also works on recent Intel CPUs with the latest hardware mitigations.
Finally, we also demonstrate the attack on CPUs previously believed to be unsusceptible to the prefetch address-translation attack, \ie ARM, IBM Power9, and AMD CPUs.

\subheading{Contributions.} The main contributions of this work are:
\begin{enumerate}[nolistsep,align=left, leftmargin=14pt, labelwidth=0pt, itemindent=!]
\item We empirically confirm the results of previous works whilst discovering an incorrect attribution of the root cause~\cite{Xiao2019Speechminer,Gruss2016Prefetch,Lipp2018meltdown}.
\item We show that the underlying root cause is speculative execution. Therefore, CPUs from other hardware vendors like AMD, ARM, and IBM are also affected.
Furthermore, the effect can even be triggered from JavaScript.
\item We discover a novel way to exploit speculative dereferences, enabling direct leakage of data values stored in registers.
\item We show that this effect, responsible for Meltdown from non-L1 data, can be adapted to Foreshadow by using addresses not valid in any address space of the guest.
\item We analyze the implications for Meltdown and Foreshadow attacks and show that Foreshadow attacks on data from the L3 cache are possible, even with Foreshadow mitigations enabled, when the unrelated Spectre-BTB mitigations are disabled.
\end{enumerate}

\subheading{Outline.}
The remainder of the paper is organized as follows.
In \cref{sec:background}, we provide background on virtual memory, cache attacks, and transient-execution attacks.
In \cref{sec:analysis}, we analyze the underlying root cause of the observed effect.
In \cref{sec:comparison}, we demonstrate the same effect on different architectures and improve the leakage rate.
In \cref{sec:ccc}, we measure the capacity using a covert channel.
In \cref{sec:vm}, we demonstrate an attack from a virtual machine.
In \cref{sec:value}, we leak actual data with seemingly harmless prefetch gadgets.
In \cref{sec:js}, we present a JavaScript-based attack leaking physical and virtual address information.
In \cref{sec:discussion}, we discuss the implications of our attacks.
We conclude in \cref{sec:conclusion}.


\section{Background and Related Work}\label{sec:background}
In this section, we provide a basic introduction to address translation, CPU caches, cache attacks, Intel SGX, and transient execution.
We also introduce transient-execution attacks and defenses.

\subsection{Address Translation}\label{sec:translation}

Virtual memory is a cornerstone of today's system-level isolation.
Each process has its own virtual memory space and cannot access memory outside of it.
In particular, processes cannot access arbitrary physical memory addresses.
The KAISER patch~\cite{Gruss2017KASLR} introduces a strong isolation between user-space and address space, meaning that kernel memory is not mapped when running in user-space.
Before the KAISER technique was applied, the virtual address space of a user process was divided into the user and kernel space.
The user address space was mapped as user-accessible while the kernel space was only accessible when the CPU was running in kernel mode.
While the user's virtual address space looks different in every process, the kernel address space looks mostly identical in all processes.
To switch from user mode to kernel mode, the x86\_64 hardware requires that parts of the kernel are mapped into the virtual address space of the process.
When a user thread performs a syscall or handles an interrupt, the hardware simply switches into kernel mode and continues operating in the same address space.
The difference is that the privileged bit of the CPU is set, and kernel code is executed instead of the user code.
Thus, the entire user and kernel address mappings remain generally unchanged while operating in kernel mode.
As sandboxed processes also use a regular virtual address space that is primarily organized by the kernel, the kernel address space is also mapped in an inaccessible way in sandboxed processes.

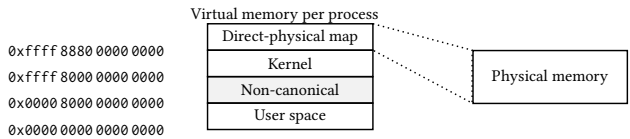
\begin{figure}[t]
    \centering
\resizebox{\hsize}{!}{
    \input{virt_to_phys.tikz}
}
\caption{Physical memory is mapped into the huge virtual address space.}
    \label{fig:direct_mapping}
\end{figure}

Many operating systems map physical memory directly into the kernel address space~\cite{Kernel2009,Levin2012}, as shown in~\Cref{fig:direct_mapping}, \eg to access paging structures and other data in physical memory.
Para-virtualizing hypervisors also employ a direct map of physical memory~\cite{Xen2009}.
Thus, every user page is mapped at least twice: once in user space and once in the kernel direct map.
When performing operations on either one of the two virtual addresses, the CPU translates the corresponding address to the same physical address.
The CPU then performs the operation based on the physical address.

For security reasons, access to virtual-to-physical address information requires root privileges~\cite{linux_patch_pagemap}.
The address-translation attack described in the Prefetch Side-Channel Attacks paper~\cite{Gruss2016Prefetch} obtains the physical address for any virtual address mapped in user space without root privileges.
For the sake of brevity, we do not discuss the translation-level oracle also described in the Prefetch Side-Channel Attacks paper~\cite{Gruss2016Prefetch} which is an orthogonal attack and, to the best of our knowledge, works as described in the paper.

\subsection{CPU Caches}\label{sec:caches}
Modern CPUs have multiple cache levels, hiding latency by buffering slower memory levels.
Page tables are stored in memory and thus are cached by the regular data caches~\cite{Intel_vol3}.
Page translation data is also stored in dedicated caches, called translation-lookaside buffers (TLBs), to speed up address translation.
Software prefetch instructions hint to the CPU that a memory address will soon be accessed in execution and so it should be fetched into the cache early to improve performance.
However, the CPU can ignore these hints~\cite{Intel_opt}.
Intel and AMD x86 CPUs have five software prefetch instructions:
\texttt{prefetcht0}, \texttt{prefetcht1}, \texttt{prefetcht2}, \texttt{prefetchnta}, \texttt{prefetchw}, and on some models the \texttt{prefetchwt1}.
On ARMv8-A CPUs we can instead use the \texttt{prfm} instruction and on IBM Power9 the \texttt{dcbt} instruction.

\subsection{Cache Attacks}\label{sec:cacheattacks}
Cache attacks have been studied for more than two decades~\cite{Kocher1996,Page2002,Tsunoo2003,Bernstein2005,Percival2005,Osvik2006}.
Today, most attacks use either \PrimeProbe~\cite{Osvik2006}, where an attacker occupies parts of the cache and waits for eviction due to cache contention with the victim, or \FlushReload~\cite{Yarom2014Flush}, where an attacker removes specific (read-only) shared memory from the cache and waits for a victim process to reload it.
\PrimeProbe has been used for many powerful cross-core covert channels and attacks~\cite{Ristenpart2009,Zhang2011,Liu2015Last,Oren2015,Lipp2016,Maurice2017Hello,Schwarz2017MGX}.
\FlushReload requires shared (read-only) memory, but is more accurate and thus has been the technique of choice in local cross-core attacks~\cite{Gruss2015Template,Guelmezoglu2015,Zhang2014,Irazoqui2015Neighbor,Irazoqui2015Lucky}.
\FlushReload has been used as a more generic primitive to test whether a memory address is in the cache or not~\cite{Lipp2018meltdown,Kocher2019,Vanbulck2018foreshadow,Schwarz2018DF}.

\subparagraph{\textbf{Prefetching attacks.}}
Gruss~\etal\cite{Gruss2016Prefetch} observed that software prefetches appear to succeed on inaccessible memory.
Using this effect on the kernel direct-physical map enables the user to fetch arbitrary physical memory into the cache.
The attacker guesses the physical address for a user-space address, tries to prefetch the corresponding address in the kernel's direct-physical map, and then uses \FlushReload on the user-space address.
If \FlushReload observes a hit, then the guess was correct.
Hence, the attacker can determine the exact physical address for any virtual address, re-enabling side-channel~\cite{Maurice2017Hello,Pessl2016} and Rowhammer attacks~\cite{Seaborn2015BH,Kim2014}.

\subsection{Intel SGX}
Intel SGX is a trusted execution mechanism enabling the execution of trusted code in a separate protected area called an enclave.
This feature was introduced with the Skylake microarchitecture as an instruction-set extension~\cite{Intel_vol3}.
The hardware prevents access to the code or data of the enclave from any source other than the enclave code itself~\cite{Intel_SGX2}.
All code running outside of the enclave is treated as untrusted in SGX.
Thus, code containing sensitive data is protected in the enclave even if the host operating system or hypervisor is compromised.
Enclave memory is mapped in the virtual address space of the host application but is inaccessible to the host. The enclave has full access to the virtual address space of its host application to share data between enclave and host.
However, as has been shown in the past, it is possible to exploit SGX via memory corruption~\cite{Lee2017SGXROP,Schwarz2018DF}, ransomware~\cite{Schwarz2019SGXMalware}, side-channel attacks~\cite{Brasser2017sgx,Schwarz2017MGX}, and transient-execution attacks~\cite{Vanbulck2018foreshadow,Schwarz2019ZL,VanSchaik2019RIDL}.

\subsection{Transient Execution}
Modern CPUs split instructions into micro-ope\-rations (\muops)~\cite{Fog2016}.
The \muops can be executed \textit{out of order} to improve performance and later on retire \textit{in order} from reorder buffers.
However, the \textit{out-of-order} stream of \muops is typically not linear.
There are branches which determine which instructions, and thereby \muops, follow next.
This is not only the case for control-flow dependencies but also data-flow dependencies.
As a performance optimization, modern CPUs rely on prediction mechanisms which predict which direction should be taken or what the condition value will be.
The CPU then speculatively continues based on its control-flow or data-flow prediction.
If the prediction was correct, the CPU utilized its resources more efficiently and saved time.
Otherwise, the results of the executed instructions are discarded, and the architecturally correct path is executed instead.
This technique is called speculative execution.
Intel CPUs have multiple branch prediction mechanisms~\cite{Intel_opt}, including the Branch History Buffer (BHB)~\cite{Bhattacharya2017perf,Kocher2019}, Branch Target Buffer (BTB)~\cite{Lee2017Inferring,Evtyushkin2016ASLR,Kocher2019}, Pattern History Table (PHT)~\cite{Fog2016,Kocher2019}, and Return Stack Buffer (RSB)~\cite{Fog2016,Maisuradze2018spectre5,Koruyeh2018spectre5}.
Lipp~\etal\cite{Lipp2018meltdown} defined instructions executed out-of-order or speculatively but not architecturally as \textit{transient instructions}.
These \textit{transient instructions} can have measurable side effects, \eg modification of TLB and cache state.
In transient-execution attacks, these side effects are then measured.

\subsection{Transient-Execution Attacks \& Defenses}
As transient execution can leave traces in the microarchitectural state, attackers can exploit these state changes to extract sensitive information.
This class of attacks is known as transient-execution attacks~\cite{Canella2019A,Intel2020Refined}.
In Meltdown-type attacks~\cite{Lipp2018meltdown} an attacker deliberately accesses memory across isolation boundaries, which is possible due to deferred permission checks in out-of-order execution.
Spectre-type attacks~\cite{Kocher2019,Kiriansky2018speculative,Chen2018SGXpectre,Horn2018spectre4,Koruyeh2018spectre5,Maisuradze2018spectre5,Schwarz2019netspectre} exploit misspeculation in a victim context.
The attacker may facilitate this misspeculation, \eg by mistraining branch predictors.
By executing along the misspeculated path, the victim inadvertently leaks information to the attacker.
To mitigate Spectre-type attacks several mitigations were developed~\cite{IntelMitigations}.
For instance, retpoline~\cite{Intel2018retpoline} replaces indirect jump instructions with \texttt{ret} instructions.
Therefore, the speculative execution path of the \texttt{ret} instruction is fixed to a certain path (e.g. to an endless loop) and does not misspeculate on potential code paths that contain Spectre gadgets.
Foreshadow~\cite{Vanbulck2018foreshadow} is a Meltdown-type attack exploiting a cleared present bit in the page table-entry.
It only works on data in the L1 cache or the line fill buffer~\cite{VanSchaik2019RIDL,Schwarz2019ZL}, which means that the data must have been recently accessed prior to the attack.
An attacker cannot directly access the targeted data from the Foreshadow attack context, and hence a widely accepted mitigation is to flush the L1 caches and line fill buffers upon context switches and to disable hyperthreading~\cite{Intel2018white_paper}.


\section{From Address-translation Attack to Foreshadow-L3}\label{sec:analysis}
In this section, we systematically analyze the properties of the address-translation attack that were erroneously explained to be caused by the insecure behavior of software prefetch instructions.\footnote{This attack is detailed in Section 3.3 and Section 5 of the Prefetch Side-Channel Attacks paper~\cite{Gruss2016Prefetch}. It should not be confused with the translation-level oracle described in Section 3.2 and Section 4 of that paper~\cite{Gruss2016Prefetch}, which to the best of our knowledge has a correct technical explanation. We focus on the part that the authors confirmed to be incorrect, \ie the address-translation attack in Section 3.3 and Section 5.}
We show that the address-translation attack~\cite{Gruss2016Prefetch} originally motivating the KAISER technique~\cite{Gruss2017KASLR} was never related to prefetch instructions.
Instead, it exploits a Spectre-BTB gadget~\cite{Canella2019A} in the kernel and, as such, is not mitigated by the KAISER technique.\footnote{This was also independently confirmed by authors of the Prefetch Side-Channel Attacks paper~\cite{Gruss2016Prefetch} that are not co-authors of this paper.}

In the address-translation attack~\cite{Gruss2016Prefetch} the attacker tries to verify whether two virtual addresses $p$ and $\bar{p}$ map to the same physical address.
For instance, on Linux, the corresponding direct-physical map address in the kernel can be used to verify the mapping.
The attacker first flushes the user-space virtual address $p$.
Then, the inaccessible (direct physical map address) $\bar{p}$ is prefetched using a software prefetch instruction.
The address $p$ is reloaded, and the timing of the reload is checked to verify whether the address is cached or uncached.
If a cache hit is observed, the inaccessible virtual address $\bar{p}$ maps to the same physical address as the virtual address $p$.
This procedure of flushing and reloading a virtual address is referred to as \FlushReload~\cite{Yarom2014Flush}.
The \FlushReload part of the address-translation attack has an F1-Score very close to 1~\cite{Yarom2014Flush}, meaning that if there is a cache hit, it will be observed in virtually every case.
The limiting factor of the attack is the probability that the guessed address is successfully ``prefetched'', as not every ``prefetch'' attempt brings the target address into the cache.
Hence, we measure the attack performance in successful \emph{fetches per second}.
More fetches per second means a shorter time to mount an attack, \eg one successful cache fetch enables leakage of 64 bytes in a Foreshadow attack, despite Foreshadow mitigations being enabled.

The prefetching component of the original attack's proof-of-concept implementation~\cite{IAIK2016Github} is shown in \Cref{lst:prefetchcpp}.
The compiled and disassembled code can be found in~\cref{lst:prefetchdisasm}.
We analyze the original attack and observe the following requirements are described for the address-translation attack to succeed:
\begin{enumerate}[nolistsep,align=left, leftmargin=17pt, labelwidth=0pt, itemindent=!]
\item[\textbf{H1}] the \texttt{prefetch} instruction (to instruct the prefetcher to prefetch);\footnote{``Our attacks are based on weaknesses in the hardware design of prefetch instructions''~\cite{Gruss2016Prefetch}.}
\item[\textbf{H2}] the value stored in the register used by the \texttt{prefetch} instruction (to indicate which address the prefetcher should prefetch);\footnote{``2. Prefetch (inaccessible) address $\bar{p}$. 3. Reload $p$. [...] the \emph{prefetch of $\bar{p}$ in step 2 leads to a cache hit} in step 3 with a high probability.''~\cite{Gruss2016Prefetch} with emphasis added.}
\item[\textbf{H3}] the \texttt{sched\_yield} syscall (to give time to the prefetcher);\footnote{``[...] delays were introduced to lower the pressure on the prefetcher.''~\cite{Gruss2016Prefetch}. These delays were implemented using a different number of \texttt{sched\_yield} system calls, as can also be seen in the original attack code~\cite{IAIK2016Github}.}
\item[\textbf{H4}] the use of the \texttt{userspace\_accessible} bit (as kernel addresses could otherwise not be translated in a user context);\footnote{``Prefetch can fetch inaccessible privileged memory into various caches on Intel x86.''~\cite{Gruss2016Prefetch} and corresponding NaCl results.}
\item[\textbf{H5}] an Intel CPU -- the ``prefetching'' effect only occurs on Intel CPUs, and other CPU vendors are not affected.\footnote{``[...] we were not able to build an address-translation oracle on [ARM] Android. As the prefetch instructions do not prefetch kernel addresses [...]''~\cite{Gruss2016Prefetch} describing why it does not work on ARM-based Android devices.}
\end{enumerate}
We test each of the above hypotheses in this section.

\begin{listing}[t!]
\begin{lstlisting}[language=C++,style=customc]
for (size_t i = 0; i < 3; ++i) {
  sched_yield();
  prefetch(direct_phys_map_addr);
}
\end{lstlisting}
\vspace{-0.4cm}
\caption[Original code]{Original code of the released proof-of-concept implementation for the address-translation attack~\cite{IAIK2016Github} from Gruss~\etal\cite{Gruss2016Prefetch}.\footnotemark
The code ``prefetches'' a (guessed) physical address from the direct physical map.
If the ``prefetch'' was successful and the physical address guess correct, the attacker subsequently observes a cache hit on the corresponding user-space address.}
\label{lst:prefetchcpp}
\end{listing}
\footnotetext{This attack is detailed in Section 3.3 and Section 5 of the Prefetch Side-Channel Attacks paper~\cite{Gruss2016Prefetch} and should not be confused with the translation-level oracle described in Section 3.2 and Section 4 of the Prefetch Side-Channel Attacks paper~\cite{Gruss2016Prefetch}.}

\begin{listing}[t!]
\begin{lstlisting}[language={[x64]Assembler},style=customasm]
; %r14 contains the direct physical address
12b6: e8 c5 fd ffff callq  1080 <sched_yield@plt>
12bb: 41 0f 18 06     prefetchnta (%r14)
12bf: 41 0f 18 1e     prefetcht2 (%r14)
12c3: e8 b8 fd ffff callq  1080 <sched_yield@plt>
12c8: 41 0f 18 06     prefetchnta (%r14)
12cc: 41 0f 18 1e     prefetcht2 (%r14)
12d0: e8 ab fd ffff callq  1080 <sched_yield@plt>
12d5: 41 0f 18 06     prefetchnta (%r14)
12d9: 41 0f 18 1e     prefetcht2 (%r14)
\end{lstlisting}
\vspace{-0.4cm}
\caption{Disassembly of the prefetching component of the prefetch address-translation attack.}
\label{lst:prefetchdisasm}
\end{listing}

\begin{listing}[t!]
\begin{lstlisting}[language={[x64]Assembler},style=customasm]
; %r14 contains the direct physical address
12b6: e8 c5 fd ffff callq  1080 <sched_yield@plt>
12bb: 0f 1f 40 00     nop
12bf: 0f 1f 40 00     nop
12c3: e8 b8 fd ffff callq  1080 <sched_yield@plt>
12c8: 0f 1f 40 00     nop
12cc: 0f 1f 40 00     nop
12d0: e8 ab fd ffff callq  1080 <sched_yield@plt>
12d5: 0f 1f 40 00     nop
12d9: 0f 1f 40 00     nop
\end{lstlisting}
\vspace{-0.4cm}
\caption{The \texttt{prefetch} instructions in the address-translation attack are replaced by 4-byte \texttt{nop}s.}
\label{lst:prefetchdisasmnop}
\end{listing}

\subsection{H1: Prefetch instruction required}
The first hypothesis is that the \texttt{prefetch} instruction is necessary for the address-translation attack.
The reasoning is that the instruction causes the prefetcher to start prefetching the provided address even though the permission check for this address fails.
To test this hypothesis, we replaced the \texttt{prefetch} instructions with \texttt{nop} instructions of the same length, as shown in~\cref{lst:prefetchdisasmnop}.
Surprisingly, the empirical result for this modified attack is identical to the original attack: there is no change in the number of cache fetches, even though there is no \texttt{prefetch} instruction in the code.
In both cases, approx. $60$ cache fetches per second occur (on an i7-8700K, Ubuntu 18.10 with kernel \texttt{4.15.0-55})\footnote{We used the original code from GitHub for comparison~\cite{IAIK2016Github} that was used to generate Figure 6 in their paper~\cite{Gruss2016Prefetch}.}
Hence, as the empirical result for the address-translation attack does not change with or without the \texttt{prefetch} instruction, we conclude that the \texttt{prefetch} instruction is not a requirement for the address-translation attack.\footnote{To the best of our knowledge, it is required for the other attack, \ie the translation-level oracle, presented by Gruss~\etal\cite{Gruss2016Prefetch}.}

\subsection{H2: Values in registers required}
The second hypothesis is that providing the direct-physical map address via the register is necessary.
We reproduced the results from Gruss~\etal\cite{Gruss2016Prefetch}, \ie that a virtual address stored in the register is the one fetched into the cache in the address-translation attack.

While we already excluded software prefetching as the root cause, the original code (\cf \cref{lst:prefetchcpp} and the modified attack code from~\cref{lst:prefetchdisasmnop}) could, in fact, trigger a hardware prefetcher.
There are patents describing CPUs that train a predictor whenever a register value is dereferenced to prefetch memory locations pointed to by register values ahead of time in subsequent runs, reducing instruction latency~\cite{IntelAdaptivePrefechting2016}.
We disable all hardware prefetchers via the model-specific register \texttt{0x1a4}~\cite{Intel_DisableHWPrefetcher} and rerun the experiment from \texttt{H1}.
In this experiment, we still observe approx. $60$ cache fetches per second, \ie disabling the prefetchers has no effect.
Hence, this already rules out any of the documented prefetchers as the root cause.

We run the modified address-translation attack uninterrupted and without context switches (and without \texttt{sched\_yield}) on one core.
In this experiment, we do not observe any cache fetches on our i7-8700K with Linux 4.15.0-55 when running this address-translation attack for $10$ hours on an isolated core (\ie no interrupts).
Hence, we conclude that it is not pure register loading that triggers the effect.
Still, the value in the register influences what is fetched into the cache.

The registers that must be used vary across kernel versions.\footnote{The original paper describes that ``delays were introduced to lower the pressure on the prefetcher''~\cite{Gruss2016Prefetch}. In fact, this was done via recompilation. Note that recompilation with additional code inserted may have side effects such as a different register allocation, that we analyze in this subsection.}
On Ubuntu 18.10 (kernel \texttt{4.18.0-17}), we observe cache hits if the registers \texttt{r12,r13} and \texttt{r14} are filled.
If we omit these registers, we do not observe any cache hits.
On Debian 8 (kernel \texttt{4.19.28-2} and Kali Linux \texttt{5.3.9-1kali1}), the registers \texttt{r9} and \texttt{r10} cause the leakage and on Linux Mint 19 (kernel \texttt{4.15.0-52}) \texttt{rdi} and \texttt{rdx} cause the leakage.
Regardless of the kernel version, we observe many cache hits when prefetching a user-space address via instruction-pointer-relative addressing, \ie the virtual address to prefetch is never in a register.
However, there is no cache hit if we use an instruction-pointer-relative address pointing into the kernel address space. 
Similarly, when specifying the target address using an x86 complex addressing mode, we only see prefetches for user-space addresses but not for kernel-space addresses. 
We only confirmed leakage if absolute virtual addresses are placed in registers. 

We developed a variant of the address-translation attack, which loads the address into most of the general-purpose registers.
This variant consistently works across all Linux versions, even with KAISER enabled.
Thus, the KAISER technique never protected against this attack.
Instead, the implementation merely changed the required registers, mitigating only the specific attack implementation and attack binary.
On an Intel Xeon Silver 4208 CPU, which has in-silicon patches against Meltdown~\cite{Lipp2018meltdown}, Foreshadow,~\cite{Vanbulck2018foreshadow} and ZombieLoad~\cite{Schwarz2019ZL}, we still observe about $30$ cache fetches per second on Ubuntu 19.04 (kernel \texttt{5.0.0-25}).

On Windows 10 (build \texttt{1803.17134}), there is no direct physical mapping we can use to fetch addresses into the cache and verify the mapping.
We fill all general-purpose registers with a kernel address and perform the syscall \texttt{SwitchToThread}.
Afterwards, we perform \FlushReload in a kernel driver to verify the speculative dereferencing in the kernel.
We observe about $15$ cache fetches per second for our kernel address.

\subsection{H3: sched\_yield required}\label{sec:h3}
The third hypothesis is that the \texttt{sched\_yield} syscall is required for the address-translation attack to work.

The idea is that for the prefetcher to consider our prefetching hint it must not be under high pressure already.
We observed in the previous experiment that omitting the \texttt{sched\_yield} syscall causes the address-translation attack to fail.
Hence, we run the experiment with no \texttt{sched\_yield} syscalls but with a large number of context switches using interrupts, \eg by running \texttt{stress -i} or \texttt{stress -d}.
Our results show that there is indeed another source of leakage resulting in cache fetches: whilst syscall handling is a primary source of leakage, further leakage occurs due to either context switching or handling of interrupts.

We first investigate whether the \texttt{sched\_yield} in the address-translation attack can be replaced by other syscalls.
We discover that other syscalls \eg \texttt{gettid}, \texttt{pipe}, \texttt{write}, expose a similar number of cache fetches.
This shows that \texttt{sched\_yield} can be replaced with arbitrary syscalls.

We then investigate whether there might be another leakage source, in particular whether context switches or interrupts trigger leakage.
We create another experiment where one process fills the registers with a chosen address in a loop, but never performs a syscall.
Another process runs \FlushReload in a loop on this specific address.
We observe about $15$ cache fetches per second on this address if the process filling the registers gets interrupted continuously, \eg due to NVMe interrupts, keystrokes, window events, or mouse movement.

These hits appear to be similarly caused by speculative execution in the interrupt handler.
Hence, we conclude that the essential part is performing syscalls or interrupts while specific registers are filled with an attacker-chosen address.

\subsection{H4: userspace\_accessible bit required}\label{sec:h4}
The fourth hypothesis is that user-mapped kernel pages are required, \ie access is prevented via the \texttt{userspace\_accessible} bit.

We constructed an experiment where we allocate several pages of memory with \texttt{mmap}.
Cache lines $A$ and $B$ are on different pages in this \texttt{mmap}'d region.
The loop (in user space) dereferences $A$ and then reloads and flushes it to see whether it was cached in each loop iteration.
In the last loop iteration only, we speculatively exchange the register value $A$ with either the address of $B$ or the direct-physical map address of $B$.
Hence, both the architectural and speculative dereferences happen at the same instruction pointer value and in the same register.
If we are training a hardware prefetcher based on the register values, we can expect it to prefetch $B$ into the cache in the speculative run.
When dereferencing $B$ directly, it is usually cached after the loop when the direct-physical map address of $B$ is used.
However, when we dereference $A$ with its value speculatively exchanged for either the address of $B$ or the direct-physical map address of $B$, $B$ is never cached after the final run.

When disabling interrupts, we observe no cache hits on $B$ on an Intel i7-4760HQ, i7-8700K, and an AMD Phenom II 1090t.
As a null hypothesis test, we perform the same test but also access $A$ in the last round.
We then should not see any cache hits on address $B$.
And indeed, none of our CPUs fetched $B$ into the cache in this scenario.

We constructed a second experiment to confirm whether the root cause of the ``prefetching'' effect lies in the user or kernel space.
While the original address-translation attack fetches addresses in the kernel direct-physical map, we can also try to fetch user addresses.
However, we discovered that this only works when SMAP is disabled (using \texttt{nosmap} kernel boot flag).
Thus, the root cause of the address-translation attack is a mechanism that adheres to SMAP (supervisor-mode access prevention) and is rooted in the kernel.
This also correlates with the finding of Kocher~\etal\cite{Kocher2019} that speculative execution cannot bypass SMAP.
Hence, we can conclude that the root cause is some form of code execution in the kernel.

\subsection{H5: Effect only on Intel CPUs}
The fifth hypothesis is that the ``prefetching'' effect only occurs on Intel CPUs.
We assume that all types of CPUs vulnerable to Spectre are also affected by the speculative dereferencing in the kernel~\cite{Kocher2019}.

Thus, we evaluate the same experiment explained in \Cref{sec:h4} on an AMD Ryzen Threadripper 1920X (Ubuntu 17.10, \texttt{4.13.0-46\-generic}), an ARM Cortex-A57 (Ubuntu 16.04.6 LTS, \texttt{4.4.38-tegra}) and an IBM Power9 (Ubuntu 18.04, \texttt{4.15.0-29}).
On the AMD Ryzen Threadripper 1920X, we achieve up to $20$ speculative fetches per second.
There, we observed a cache hit rate of $0.0004\%$ on $B$, which is the standard false positive rate we observed for \FlushReload attacks on this CPU.
On the Cortex-A57, we observe $5$ speculative fetches per second, and on the IBM Power9, we detect up to $15$ speculative fetches per second.
We do not observe any false positives on the ARM and Power9 CPUs during this experiment.

We run the same experiment on a Raspberry Pi 3 (ARM Cortex-A53,Ubuntu 18.04, kernel \texttt{4.15.0}), an in-order CPU with no branch prediction~\cite{ARMCortexA53}.
Thus, this CPU is not susceptible to any Spectre-type attacks.
Running the same code for $1$ hour, we do not observe any cache fetches.
Therefore, as no leakage appears on an in-order CPU without branch prediction, the effect must be related to Spectre.
The hypothesis that the effect is hardware-specific to Intel CPUs is incorrect; any CPU susceptible to Spectre-BTB is vulnerable to speculative dereferencing in the kernel if the mitigations are not enabled.

\subsection{Speculative Execution in the Kernel}\label{sec:specex}
From the previous analysis of the hypotheses, we can conclude that the leakage is not due to the software or hardware prefetchers but due to speculative code execution in the kernel. 
While this conclusion might not be suprising with the knowledge of Spectre, Spectre was only discovered one year after the original prefetch paper~\cite{Gruss2016Prefetch} was published.
We now show that the primary leakage is caused by Spectre-BTB-SA-IP (branch target buffer, training in same address space, and in-place)~\cite{Canella2019A}.

\begin{listing}[t!]
\begin{lstlisting}[language={[x64]Assembler},style=customasm]
;<do_syscall_64+106>
=> 0xffffffff8100134a: callq 0xffffffff81802000
=> 0xffffffff81802000: jmpq   *%rax
; with retpoline
=> 0xffffffff81802000:  callq  0xffffffff8180200c
=> 0xffffffff8180200c:  mov    %rax,(%rsp)
=> 0xffffffff81802010:  retq
\end{lstlisting}
\vspace{-0.4cm}
\caption{While processing a syscall, the kernel performs multiple indirect jumps, \eg one to the corresponding syscall handler.
With retpoline~\cite{Turner2018retpoline}, the kernel uses a \texttt{retq} for the indirect jump.
Without retpoline the \texttt{jmp} instruction is used on a pointer in a register.}
\label{lst:retpolinekernel}
\end{listing}

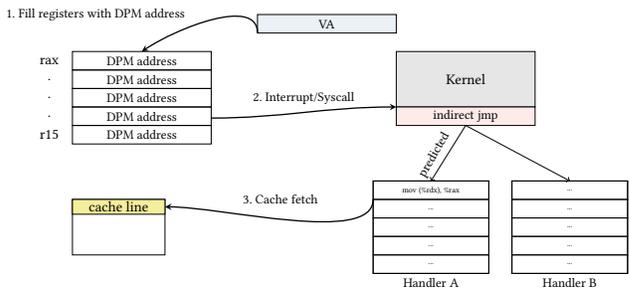
\begin{figure}[t]
 \centering
 \resizebox{\hsize}{!}{
    \input{spec_acces_in_kernel.tikz}
 }
 \caption{The kernel speculatively dereferences the direct-physical map address (DPM). With \FlushReload, we observe cache hits on the corresponding user-space address.}
 \label{fig:speculative_execution}
\end{figure}

First, we observe that during a syscall, the kernel performs multiple indirect jumps to execute the corresponding system-call handler (\cf \Cref{lst:retpolinekernel}).
With retpoline, the kernel uses a \texttt{retq} for the indirect jump, which traps the speculative execution path to a fixed branch.
Without retpoline, the \texttt{jmp} instruction is used on a pointer in a register.
This causes speculative execution based on Spectre-BTB-SA-IP.
The address-translation attack then succeeds because different syscalls use a different number of arguments.
The unified interface does not zero out registers that a given syscall does not require.
Consequently, during speculative execution, the CPU might use an incorrect prediction from the branch-target buffer (BTB) and speculate into the wrong syscall.
\cref{fig:speculative_execution} illustrates the speculative execution in the kernel dereferencing.
In this misspeculated syscall, registers containing attacker-chosen addresses are used.
This can either be because the registers were never initialized and instead still contain the attacker-chosen addresses, or because they are deliberately initialized to attacker-chosen addresses through the syscall entry code.

We evaluate the leakage rate of other syscalls and the impact of mistraining the branch prediction mechanisms in \Cref{sec:comparison}.
On recent kernels, the leakage completely disappears unless \texttt{nospectre\_v2} (\ie disable Spectre-BTB countermeasures) is passed as a boot flag.
Disabling the Spectre V2 mitigations is interesting for cloud computing since the mitigations introduce a big performance overhead~\cite{Slashdot2019RetpolinePerformance}.
Thus, the address-translation attack is mitigated using the Spectre-BTB countermeasures and not, as described in previous work~\cite{Gruss2016Prefetch,Gruss2017KASLR}, by KAISER (KPTI)~\cite{Gruss2017KASLR}, or LAZARUS~\cite{Gens2017}.

We observed other speculative execution in the kernel that exposes the same effects.
However, we observe $15$ speculative fetches per second on an i5-8250U (kernel \texttt{5.0.0-20}) if we eliminate the Spectre-BTB-SA-IP leak from \cref{lst:retpolinekernel}, empirically confirming that this is one of the main leakage sources.
As already mentioned, there are further Spectre gadgets in the interrupt handling.

As Canella~\etal\cite{Canella2019A} showed, there were about $172$ unmitigated Spectre v1 ``prefetch'' gadgets found in the Linux kernel.
These gadgets enable the same attacks as presented in this paper.
Currently there is no consistent plan to mitigate these gadgets.
However, any prefetch gadget can be used for an address-translation attack~\cite{Gruss2016Prefetch} and thus would also re-enable Foreshadow-VMM attacks~\cite{Vanbulck2018foreshadow,Weisse2018foreshadow}.
As concurrent work showed, there are gadgets in the Linux kernel which can be used to fetch data into the L1D cache in Xen~\cite{XENL1TFPrefetch} and an artificial gadget was exploited by Stecklina~\cite{StecklinaL1TFKVM}.

In the case of interrupts, we analyzed the interrupt handling in the Linux kernel version 4.19.0 and observed that the register values from \texttt{r8-r15} are cleared but stored on the stack and restored after the interrupt.
Thus, either there is a misspeculation on old register values, or the leakage comes from the stored stack values~\cite{Maisuradze2018spectre5}.
Additionally, we found several \texttt{jmp} instructions that occur in the analyzed instruction trace, which might trigger speculative cache fetches.
Again, when using the Spectre-BTB mitigations we could not detect any leakage while triggering interrupts, showing that this is a crucial element for the speculative dereferencing.

\subsection{Meltdown-L3 and Foreshadow-L3}
The speculative dereferencing was also noticed but misattributed to the prefetcher in subsequent work.
For instance, the Meltdown paper~\cite{Lipp2018meltdown} reports that data is fetched from L3 into L1 while mounting a Meltdown attack.
Van Bulck~\etal\cite{Vanbulck2018foreshadow} did not observe this prefetching effect for Foreshadow.
Based on this observation, further works also mentioned this effect without analyzing it thoroughly~\cite{Canella2019A,Nilsson2020SGXSurvey,VanSchaik2019RIDL}.
In \Speechminer the explanation provided is that performing a Meltdown-US attack causes data to be repeatedly prefetched from L1 to L3~\cite{Xiao2019Speechminer}.

We used similar Meltdown-L3 setups as \Speechminer~\cite{Xiao2019Speechminer} and Meltdown~\cite{Lipp2018meltdown}.
For this purpose, we contacted the authors to ask for their specific experiment setup.
According to the authors of \Speechminer~\cite{Xiao2019Speechminer}, the kernel boot flags \texttt{nopti}, \texttt{nokaslr} were used on kernel 4.4.0-134.
We used Ubuntu 16.04 on an Intel i7-6700K to reproduce the attack.
The authors of Meltdown used Ubuntu 16.10 (kernel 4.8.0), which at that moment of writing did not have any mitigations against Spectre at all~\cite{Lipp2018meltdown}.

We construct our Meltdown-L3 experiment as follows.
One physical core constantly accesses a secret to ensure that the value stays in the L3, as the L3 is shared across all physical cores.
On a different physical core, we run Meltdown on the direct-physical map.
On recent Linux kernels with full Spectre v2 mitigations implemented, we could not reproduce the result on the same machine with the default mitigations enabled.
With the \texttt{nospectre\_v2} flag, our Meltdown-L3 attack works again when triggering the prefetch gadget in the kernel.
Since we run Meltdown on the direct-physical map, we place the corresponding direct-physical map address in a register.
Now, when a syscall is performed, or an interrupt is triggered, the direct-physical map address is speculatively dereferenced, causing the data to be fetched into L1.

Concluding the above experiment, on Linux kernels 4.4.0-137 and 4.8, as respectively used in \Speechminer~\cite{Xiao2019Speechminer} and Meltdown~\cite{Lipp2018meltdown}, not all Spectre-BTB mitigations such as IBPB and RSB stuffing were implemented.
Thus, the Meltdown-L3 prefetching works because these mitigations are not implemented on these kernel versions~\cite{Linux2018IBPB}.
Without our new insights that the prefetching effect is caused by speculative execution, it is almost inevitable to not misdesign these experiments, inevitably leading to incomplete or incorrect observations and conclusions on Meltdown and Foreshadow and their mitigations.
We confirmed with the authors that their experiment design was not robust to our new insight and therefore lead to wrong conclusions.

\texttt{Foreshadow-L3}¸
The same prefetching effect can be used to perform Foreshadow~\cite{Vanbulck2018foreshadow}.
If a secret is present in the L3 cache and the direct-physical map address is derefenced in the hypervisor kernel, data can be fetched into the L1.
This reenables Foreshadow even with Foreshadow mitigations enabled if the unrelated Spectre-BTB mitigations are disabled.
We demonstrate this attack in KVM in~\cref{sec:vm}.

In Meltdown and Foreshadow, as in other transient-execution attacks, common implementations transmit a secret byte from the transient-execution realm via a \FlushReload cache covert channel to the architectural realm.
Most implementations transmit 1 byte of data by accessing one of 256 offsets in an array.
Several papers, including Meltdown and Foreshadow, observed a bias towards the `0' index, where a secret value of `0' is falsely reported to the attacker.
This effect was observed and explained by the zeroing of invalid loads~\cite{Lipp2018meltdown,Vanbulck2018foreshadow}.
We also tried to reproduce these results.
However, we only observed a bias towards zero on systems with hardware mitigations against Meltdown and Foreshadow, which by design return zeros in these attack scenarios~\cite{Canella2020kaslr}.
We observed no bias towards zero on other systems with the most recent software patches and software mitigations.
To transmit a value of `0' through the \FlushReload covert channel, the offset `0' is accessed, \ie the array base address.
However, the \FlushReload array base address is stored in a register during the \FlushReload loop.
Thus, the base address is speculatively dereferenced due to interrupts and the \texttt{sched\_yield} found in the \FlushReload loops in these implementations.
This indicates that the speculative dereferencing of user-space registers creates at least part of the zero bias, if not all, since the bias is no longer visible on more recent systems with full software mitigations against Spectre enabled.


\section{Improving the Leakage Rate}\label{sec:comparison}
With the knowledge that the root cause of the prefetching effect is speculative execution in the kernel, we can try to optimize the number of cache fetches.
As already observed in \cref{sec:h3}, the \texttt{sched\_yield} syscall can be replaced by an arbitrary syscall to perform the address-translation attack.
In this section, we compare different syscalls and their impact on the number of speculative cache fetches on different architectures and kernel versions.
We investigate the impact of executing additional syscalls before and after the register filling and measure their effects on the number of speculative cache fetches.

\begin{table}[t]
  \caption{Evaluated systems, their CPUs, operating systems, and kernel versions used in the syscall evaluation.}
\vspace{-0.1cm}
\setlength{\aboverulesep}{0pt}
\setlength{\belowrulesep}{0pt}
    \begin{center}
      \resizebox{\hsize}{!}{
      \begin{tabular}{ccc}
        \hline
        CPU                              & Operating System & Kernel\\
        \hline
        Intel i5-8250U                   & Linux Mint 19 & \texttt{4.15.0-52} \\
        Intel i7-8700K                   & Ubuntu 18.04  & \texttt{4.15.0-55} \\
        ARM Cortex-A57                   & Ubuntu 16.04.6 & \texttt{4.4.38-tegra}\\
        AMD Threadripper 1920X           & Ubuntu 17.10  & \texttt{4.13.0-46} \\
      \end{tabular}
    }
  \end{center}
  \label{tab:evaluated_systems}
\end{table}

\subparagraph{\textbf{Setup.}}
\Cref{tab:evaluated_systems} lists the test systems used in our experiments.
On the Intel and AMD CPUs, we disabled the Spectre-BTB mitigations using the kernel flag \texttt{nospectre\_v2}.
On the evaluated ARM CPU, Spectre-BTB mitigations are not supported by the tested firmware.
We evaluate the speculative dereferencing using different syscalls to observe whether the number of cache fetches increases.
Based on the number of correct and incorrect cache fetches of two virtual addresses, we calculate the F1-score, \ie the harmonic average of precision and recall.

When performing a syscall, the CPU might mispredict the target syscall based on the results of the BTB.
If a misprediction occurs, another syscall which dereferences the values of user-space registers might be speculatively executed.
Therefore if we perform syscalls before we fill the registers with the direct-physical map address, we might mistrain the BTB and trigger the CPU to speculatively execute the mistrained syscall.
We evaluate the mistraining of the BTB for \texttt{sched\_yield} in \cref{sec:mistraining_yield}.

We create a framework that runs the experiment from~\cref{sec:h4} with \SIx{20} different syscalls (after filling the registers) and computes the F1-score.
We perform different syscalls before filling the registers to mistrain the branch prediction.
One direct-physical-map address has a corresponding mapping to a virtual address and should trigger speculative fetches into the cache.
The other direct-physical-map address should not produce any cache hits on the same virtual address.
If there is a cache hit on the correct virtual address, we count it as a true positive.
Conversely, if there is no hit when there should have been one, we count it as a false negative.
On the second address, we count the false positives and true negatives.
For syscalls with parameters, \eg \texttt{mmap}, we set the value of all parameters to the direct-physical-map address, \ie \texttt{mmap(addr, addr, addr, addr, addr, addr)}.
We repeat this experiment \SIx{1000} times for each syscall on each system and compute the F1-Score.

\begin{table}[t]
  \setlength{\aboverulesep}{0pt}
  \setlength{\belowrulesep}{0pt}
  \caption{F1-Scores for speculative cache fetches with different syscalls on different CPU architectures.}
\vspace{-0.1cm}
\label{tab:f1_scores}
    \begin{center}
      \resizebox{\hsize}{!}{
      \begin{tabular}{cccccc}
        \hline
        Syscall                          & Syscall executed before     & i5-8250U     & i7-8700K    & Threadripper 1920X  & Cortex-A57 \\
        \hline
        \multirow{4}{*}{sched\_yield}    & None                        &  66.40\%     & 91.49\%       & 99.29\%             & 76.61\% \\
                                         & send-to                     &  56.42\%     & 4.60\%        & 52.94\%             & 44.88\% \\
                                         & geteuid                     &  46.62\%     & 1.90\%        & 63.94\%             & 48.82\% \\
                                         & stat                        &  77.37\%     & 57.44\%       & 69.28\%             & 63.57\% \\
                                         \hline
        \multirow{4}{*}{pipe}            & None                        &  100\%       & 99.35\%       & 100\%               & 100\% \\
                                         & send-to                     &  99.9\%      & 99.60\%       & 100\%               & 100\% \\
                                         & geteuid                     &  99.9\%      & 99.61\%       & 100\%               & 100\% \\
                                         & stat                        &  99.9\%      & 99.55\%       & 99.9\%              & 100\% \\
                                         \hline
        \multirow{4}{*}{read}            & None                        & 10.42\%      & 0.09\%        & 8.50\%              & 57.95\% \\
                                         & send-to                     & 14.47\%      & 21.26\%       & 1.90\%              & 78.86\% \\
                                         & geteuid                     & 15.32\%      & 56.73\%       & 2.35\%              & 73.73\% \\
                                         & stat                        & 28.32\%      & 24.07\%       & 9.70\%              & 23.32\% \\
                                         \hline
        \multirow{4}{*}{write}           & None                        & 7.69\%       & 91.24\%       & 76.46\%             & 58.95\% \\
                                         & send-to                     & 14.29\%      & 9.88\%        & 11.00\%              & 45.68\% \\
                                         & geteuid                     & 15.49\%      & 32.21\%       & 52.94\%             & 49.47\% \\
                                         & stat                        & 9.16\%       & 9.70\%        & 52.83\%              & 12.03\% \\
                                         \hline
        \multirow{4}{*}{nanosleep}       & None                        &  21.2\%      & 27.43\%       & 52.61\%             & 87.40\% \\
                                         & send-to                     &  46.59\%     & 13.43\%       & 76.23\%             & 82.83\% \\
                                         & geteuid                     &  29.93\%     & 96.05\%       & 89.62\%             & 69.63\% \\
                                         & stat                        &  59.84\%     & 99.14\%       & 89.68\%             & 77.67\% \\
      \end{tabular}
    }
  \end{center}
\end{table}

\subparagraph{\textbf{Evaluation.}}
We evaluate different syscalls for branch prediction mistraining by executing a single syscall before and after filling the registers with the target address.
\Cref{tab:f1_scores} lists the F1-scores of syscalls which achieved the highest number of cache fetches after filling registers with addresses.
The results show that the same effects occur on both AMD and ARM CPUs, with similar F1-scores.

Executing the \texttt{pipe} syscall after filling the register seems to always trigger speculative dereferencing in the kernel on each architecture.
However, this syscall has to perform many operations and takes \SI{3} to \SI{5} times longer to execute than \texttt{sched\_yield}.
On recent Linux kernels (version \texttt{5}), we observe that the number of cache fetches decreases.
This is due to a change in the implementation of the syscall handler, and thus other paths need to be executed to increase the probability of speculative dereferencing.
We observe that an additional, different, syscall executed before filling the registers also mistrains the branch prediction.
Thus, we also compare the number of cache fetches with an additional syscall added before the registers are filled.
If we add additional syscalls like \texttt{stat}, \texttt{sendto}, or \texttt{geteuid} before filling the registers, we achieve higher F1-scores in some cases.
For instance, executing the syscalls \texttt{read} and \texttt{nanosleep} after the register filling performs significantly better (up to 80\% higher F1-scores) with prior syscall mistraining.
However, as listed in~\Cref{tab:f1_scores}, not every additional syscall increases the number of cache fetches.


\section{Covert Channel}\label{sec:ccc}
For the purpose of a systematic analysis, we evaluate the capacity of our discovered information leakage by building a covert channel.
Note that while covert channels assume a colluding sender and receiver, it is considered best practice to evaluate the maximum performance of a side channel by building a covert channel.
Similar to previous works~\cite{Pessl2016,Wu2014}, our covert channel works without shared memory and across CPU cores.
The capacity of the covert channel indicates an upper bound for potential attacks where the attacker and victim are not colluding.

\begin{figure}[t]
 \centering
 \resizebox{\hsize}{!}{
    \input{covert.tikz}
 }
 \caption{The setup for the covert channel.
 The receiver allocates a page accessible through the virtual address $v$.
 The sender uses the direct-physical mapping $p$ of the page to influence the cache state.}
 \label{fig:covert-setup}
\end{figure}
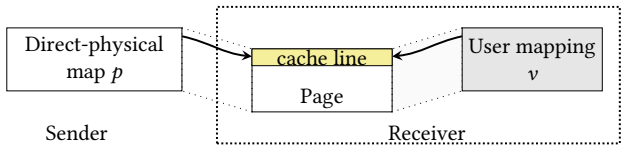

\subparagraph{\textbf{Setup.}}
\Cref{fig:covert-setup} shows the covert-channel setup.
The receiver allocates a memory page which is used for the communication.
The receiver can access the page through the virtual address $v$.
Furthermore, the receiver retrieves the direct-physical-map address $p$ of this page.
This can be done, \ie using the virtual-to-physical address-translation technique we analyzed in \cref{sec:analysis}.
The address $p$ is used by the sender to transmit data to the receiver.
The address $p$ also maps to the page, but as it is a kernel address, a user program cannot access the page via this virtual address.
The direct-physical-map address $p$ is a valid kernel address for every process.
Moreover, as the shared last-level cache is physically indexed and physically tagged, it does not matter for the cache which virtual address is used to access the page.

\subparagraph{\textbf{Transmission.}}
The transmitted data is encoded into the cache state by either caching a cache line of the receiver page (`1'-bit) or not caching the cache line of the receiver page (`0'-bit).
To cache a cache line of the receiver page, the sender uses Spectre-BTB-SA-IP in the kernel to speculatively access the kernel address $p$.
For this, the sender constantly fills all x86-64 general-purpose registers with the kernel address $p$ and performs a syscall.
The kernel address is then speculatively dereferenced in the kernel and the CPU caches the chosen cache line of the receiver page.
Hence, we can use this primitive to transmit one bit of information.
To synchronize the two processes, we define a time window per bit for sender and receiver.
On the receiver side, we reaccess the same cache line to check whether the address $v$, \ie the first cache line of the receiver page, is cached.
After the access, the receiver flushes the address $v$ to repeat the measurement.
A cache hit is interpreted as a `1'-bit.
Conversely, if the sender wants to transmit a `0'-bit, the sender does not write the value into the registers and instead waits until the time window is exceeded.
Thus, if the receiver encounters a cache miss, it is interpreted as a `0'-bit.

\subparagraph{\textbf{Evaluation.}}
We evaluated the covert channel by transmitting random messages between two processes running on different physical CPU cores.
Our test system was equipped with an Intel i7-6500U CPU, running Linux Mint 19 (kernel \texttt{4.15.0-52-generic}, \texttt{nospectre\_v2} boot flag).

In our setup, we transmit $128$ bytes from the sender to the receiver and run the experiment $50$ times.
We observed that additional interrupts on the core where the syscall is performed increases the performance of the covert channel.
These interrupts trigger the speculative execution we observed in the interrupt handler.
In particular, I/O interrupts, \ie syncing the NVMe device, create additional cache fetches.
While we achieved a transmission rate of up to \SI{30}{\bit/\second}, at this rate we had a high standard error of approx. 1\%.
We achieved the highest capacity at a transmission rate of \SI{10}{\bit/\second}.
At this rate, the standard error is, on average, 0.1\%.
This result is comparable to related work in similar scenarios~\cite{Pessl2016,Wu2014}.
To achieve an error-free transmission, error-correction techniques~\cite{Maurice2017Hello} can be used.
Compared to to the Flush+Prefetch covert channel demonstrated by Gruss~\etal\cite{Gruss2016Prefetch} is that our covert channel does not require any shared memory.
Thus, while slower, it is more powerful as it can be used in a wider range of scenarios.


\section{Speculative Dereferences and Virtual Machines}\label{sec:vm}
In this section, we examine speculative dereferencing in virtual machines.
We demonstrate a successful end-to-end attack using interrupts from a virtual-machine guest running under KVM on a Linux host~\cite{kvm_linux}.
The attack succeeds even with the recommended Foreshadow mitigations enabled, provided that the unrelated Spectre-BTB mitigations are disabled.
Against our expectations, we did not observe any speculative dereferencing of guest-controlled registers in Microsoft's Hyper-V HyperClear Foreshadow mitigation.
We provide a thorough analysis of this negative result.

\begin{figure}[t]
 \centering
 \resizebox{\hsize}{!}{
    \input{hv_l1tf.tikz}
}
\caption{If a guest-chosen address is speculatively fetched into the cache during a hypercall or interrupt and not flushed before the virtual machine is resumed, the attacker can perform a Foreshadow attack to leak the fetched data.}
\label{fig:hyperv}
\end{figure}
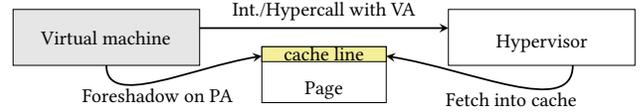

Since we observe speculative dereferencing in the syscall handling, we investigate whether hypercalls trigger a similar effect.
The attacker targets a specific host-memory location where the host virtual address and physical address are known but inaccessible.

\subparagraph{\textbf{Foreshadow Attack on Virtualization Software.}}
If an address from the host is speculatively fetched into the L1 cache on a hypercall from the guest, we expect it to have a similar speculative-dereferencing effect.
With the speculative memory access in the kernel, we can fetch arbitrary memory from L2, L3, or DRAM into the L1 cache.
Consequently, Foreshadow can be used on arbitrary memory addresses provided the L1TF mitigations in use do not flush the entire L1 data cache~\cite{UbuntuKVM2019,XENL1TFPrefetch,StecklinaL1TFKVM}.
\Cref{fig:hyperv} illustrates the attack using hypercalls or interrupts and Foreshadow.
The attacking guest loads a host virtual address into the registers used as hypercall parameters and then performs hypercalls.
If there is a prefetching gadget in the hypercall handler and the CPU misspeculates into this gadget, the host virtual address is fetched into the cache.
The attacker then performs a Foreshadow attack and leaks the value from the loaded virtual address.

\subsection{Foreshadow on Patched Linux KVM}
Concurrent work showed that prefetching gadgets in the kernel, in combination with L1TF, can be exploited on Xen and KVM~\cite{XENL1TFPrefetch,StecklinaL1TFKVM}.
The default setting on Ubuntu 19.04 (kernel \texttt{5.0.0-20}) is to only conditionally flush the L1 data cache upon VM entry via KVM~\cite{UbuntuKVM2019}, which is also the case for Kali Linux (kernel \texttt{5.3.9-1kali1}).
The L1 data cache is only flushed in nested VM entry scenarios or in situations where data from the host might be leaked.
Since Linux kernel \texttt{4.9.81}, Linux's KVM implementation clears all guest clobbered registers to prevent speculative dereferencing~\cite{kvm_register_clearing}.
In our attack, the guest fills all general-purpose registers with direct-physical-map addresses from the host.

\subparagraph{\textbf{End-To-End Foreshadow Attack via Interrupts.}}
In \Cref{sec:h3}, we observed that context switches triggered by interrupts can also cause speculative cache fetches.
We use the example from \Cref{sec:h3} to verify whether the ``prefetching'' effect can also be exploited from a virtualized environment.
In this setup, we virtualize Linux buildroot (kernel \texttt{4.16.18}) on a Kali Linux host (kernel \texttt{5.3.9-1kali1}) using qemu (\texttt{4.2.0}) with the KVM backend.
In our experiment, the guest constantly fills a register with a direct-physical-map address and performs the \texttt{sched\_yield} syscall.
We verify with \FlushReload in a loop on the corresponding host virtual address that the address is indeed cached.
Hence, we can successfully fetch arbitrary hypervisor addresses into the L1 cache on kernel versions before the patch, \ie with Foreshadow mitigations but incomplete Spectre-BTB mitigations.
We observe about $25$ speculative cache fetches per minute using NVMe interrupts on our Debian machine.
The attacker, running as a guest, can use this gadget to prefetch data into the L1.
Since data is now located in the L1, this reenables a Foreshadow attack~\cite{Vanbulck2018foreshadow}, allowing guest-to-host memory reads.
As described before, $25$ fetches per minute means that we can theoretically leak up to $64 \cdot 25 = 1600$ bytes per minute (or $26.7$ bytes per second) with a Foreshadow attack despite mitigations in place.
However, this requires a sophisticated attacker who avoids context switches once the target cache line is cached.

We develop an end-to-end Foreshadow-L3 exploit that works despite enabled Foreshadow mitigations, provided the unrelated Spectre-BTB mitigations are disabled.
In this attack the host constantly accesses a secret on a physical core, which ensures it remains in the shared L3 cache.
We assign one isolated physical core, consisting of two hyperthreads, to our virtual machine.
In the virtual machine, the attacker fills all registers on one logical core (hyperthread) and performs the Foreshadow attack on the other logical core.
Note that this is different from the original Foreshadow attack where one hyperthread is controlled by the attacker and the sibling hyperthread is used by the victim.
Our scenario is more realistic, as the attacker controls both hyperthreads, \ie both hyperthreads are in the same trust domain.
With this proof-of-concept attack implementation, we are able to leak $7$ bytes per minute successfully~\footnote{An anonymized demonstration video can be found here: https://streamable.com/8ke5ub}.
Note that this can be optimized further, as the current proof-of-concept produces context switches regardless of whether the cache line is cached or not.
Our attack clearly shows that the recommended Foreshadow mitigations alone are not sufficient to mitigate Foreshadow attacks, and Spectre-BTB mitigations must be enabled to fully mitigate our Foreshadow-L3 attack.

\subparagraph{\textbf{No Prefetching gadget in Hypercalls in KVM}}
We track the register values in hypercalls and validate whether the register values from the guest system are speculatively fetched into the cache.
We neither observe that the direct-physical-map address is still located in the registers nor that it is speculatively fetched into the cache.
However, as was shown in concurrent work~\cite{StecklinaL1TFKVM,XENL1TFPrefetch}, prefetch gadgets exist in the kernel that can be exploited to fetch data into the cache, and these gadgets can be exploited using Foreshadow.

\subsection{Negative Result: Foreshadow on Hyper-V HyperClear}
We examined whether the same attack also works on Windows 10 (build \texttt{1803.17134}), which includes the latest patch for Foreshadow.
As on Linux, we disabled the mitigations for Spectre-BTB and tried to fetch hypervisor addresses from guest systems into the cache.

Microsoft's Hyper-V HyperClear Mitigation~\cite{Microsoft2018HyperV} for Foreshadow claims to only flush the L1 data cache when switching between virtual cores.
Hence, it should be susceptible to the same basic attack we described at the beginning of this section.
For our experiment, the attacker passes a known virtual address of a secret variable from the host operating system for all parameters of a hypercall.
However, we could not find any exploitable timing difference after switching from the guest to the hypervisor.
Our experiments concerning this negative result are discussed in~\cref{sec:appendix}.


\section{Leaking Values from SGX Registers}\label{sec:value}
In this section, we present a novel method, \RegAttack, to leak register contents from an SGX enclave in the presence of only a speculative register dereference.
We show that this technique can also be generalized and applied to other contexts.
Leaking the values of registers is useful, \eg to extract parts of keys or intermediate values from cryptographic operations.
While there are already Spectre attacks on SGX enclaves~\cite{Chen2018SGXpectre,OKeeffe18sgxspectre}, they require the typical Spectre-PHT gadget~\cite{Kocher2019}, \ie a double indirect memory access after a conditional branch.

\subsection{Dereference Trap}
For \RegAttack, we exploit transient code paths inside an enclave which speculatively dereference a register containing a secret value.
The setup is similar to the kernel case we examined in \Cref{sec:specex}.
An SGX enclave has access to the entire virtual address space~\cite{Intel_SGX2}.
Hence, any speculative memory access to a valid virtual address caches the data at this address.

The basic idea of \RegAttack is to ensure that the entire virtual address space of the application is mapped.
Thus, if a register containing a secret is speculatively dereferenced, the corresponding virtual address is cached.
The attacker can detect which virtual address is cached and infer the secret.
However, in practice, there are two main challenges which must be resolved to implement \RegAttack.
Firstly, the virtual address space is much larger than the physical address space.
Thus it is not possible to simply map all virtual addresses to physical addresses.
Secondly, the \FlushReload attack is a bottleneck, as even a highly-optimized \FlushReload attack takes around 300 CPU cycles~\cite{Schwarz2018DF}.
Hence, probing every cache line of the entire user-accessible virtual address space of $2^{47}$ bytes would require around 2 days on a \SI{4}{GHz} CPU.
Moreover, probing this many cache lines does not work as the cached address does not remain in the cache if many other addresses are accessed.

\begin{figure}[t]
    \centering
\resizebox{\hsize}{!}{
    \input{value_leak.tikz}
}
\vspace{-0.2cm}
\caption{Leaking the value of an x86 general-purpose register using \RegAttack and \FlushReload on two different physical addresses. $v_0$ to $v_{n-1}$ represent the memory mappings on one of the shared memory regions.}
    \label{fig:valueleak}
\end{figure}
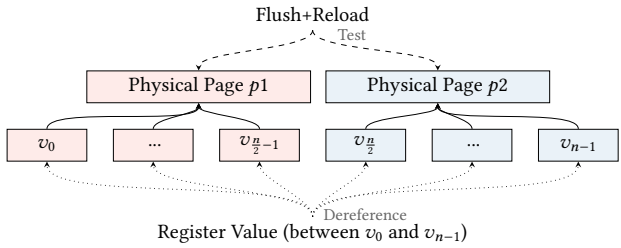

\subparagraph{\textbf{Divide and Conquer.}}
Instead of mapping every page in the virtual address space to its own physical pages, we only map 2 physical pages $p1$ and $p2$, as illustrated in \Cref{fig:valueleak}.
By leveraging shared memory, we can map one physical page multiple times into the virtual address space.
By default, the number of \texttt{mmap}ed segments which can be mapped simultaneously is limited to \SIx{65536}~\cite{max_mmap}.
However, as the attacker in the SGX threat model is privileged~\cite{Intel_SGX2} we can easily disable this limit.
The maximum allowed value is $2^{31}-1$, which makes it possible to map 1/16$^{th}$ of the user-accessible virtual address space.
If we only consider 32-bit secrets, \ie secrets which are stored in the lower half of 64-bit registers, $2^{20}$ mappings are sufficient.
Out of these, the first $2^{10}$ virtual addresses map to physical page $p1$ and the second $2^{10}$ addresses map to page $p2$.
Consequently the majority of 32-bit values are now valid addresses that either map to $p1$ or $p2$.
Thus, after a 32-bit secret is speculatively dereferenced inside the enclave, the attacker only needs to probe the 64 cache lines of each of the two physical pages.
A cache hit reveals the most-significant bit (bit 31) of the secret as well as bits 6 to 11, which define the cache-line offset on the page.

To learn the remaining bits 12 to 30, we continue in a fashion akin to binary-search.
We unmap all mappings to $p1$ and $p2$ and create half as many mappings as before.
Again, half of the new mappings map to $p1$ and half of the new mappings map to $p2$.
From a cache hit in this setup, we can again learn one bit of the secret.
We can repeat these steps until all bits from bit 6 to 31 of the secret are known.
As the granularity of \FlushReload is one cache line, we cannot leak the least-significant 6 bits of the secret.

As a privileged attacker, we can also disable the hardware prefetchers on Intel CPUs by setting the model-specific register \texttt{0x1a4} to 15~\cite{Intel_DisableHWPrefetcher}.
This prevents spurious cache hits, which is especially important for probing the cache lines on a single page.

We evaluated \RegAttack on our test system and recovered a 32-bit value stored in a 64-bit register within $15$ minutes.

\subsection{Speculative Type Confusion}
SGX registers are invisible to the kernel and can thus not be speculatively dereferenced from outside SGX.
Hence, the dereference gadget has to be inside the enclave.
While there is a mechanism similar to a context switch when an enclave is interrupted, we could not find such a gadget in either the current SGX SDK or driver code.
This is unsurprising, as this code is hardened with memory fences for nearly all memory loads to prevent LVI~\cite{Vanbulck2020lvi} as well as other transient-execution attacks.

Hence, to leak secret registers using \RegAttack, the gadget must be in the enclave code.
Such a gadget can easily be introduced, \eg when using polymorphism in C++.
\Cref{lst:specconfuse} shows a minimal example of introducing such a gadget.

\begin{listing}[t!]
\begin{lstlisting}[language=C++,escapechar=|]
class Object {
public:
  virtual void print() = 0;
};
class Dummy : public Object {
private:
  char* data;
public:
  Dummy() { data = "TEST"; }
  virtual void print() { puts(data); }
};
class Secret : public Object {
private:
  size_t secret;
public:
  Secret() { secret = 0x12300000; } |\label{line:secret}|
  virtual void print() { }
};
void printObject(Object* o) { o->print(); } |\label{line:call}|
\end{lstlisting}
\vspace{-0.4cm}
\caption{Speculative type confusion which leaks the secret of \texttt{Secret} class instances using \RegAttack.}
\label{lst:specconfuse}
\end{listing}

The virtual functions are implemented using \textit{vtable}s for which the compiler emits an indirect call in \Cref{line:call}.
The branch predictor for this indirect call learns the last call target.
Thus, if the call target changes because the type of the object is different, speculative execution still executes the function of the last object with the data of the current object.

In this code, calling \texttt{printObject} first with an instance of \texttt{Dummy} mistrains the branch predictor to call \texttt{Dummy::print}, dereferencing the first member of the class.
A subsequent call to \texttt{printObject} with an instance of \texttt{Secret} leads to speculative execution of \texttt{Dummy::print}.
However, the dereferenced member is now the secret (\Cref{line:secret}) of the \texttt{Secret} class.

The speculative type confusion in such a code construct leads to a speculative dereference of a value which would never be dereferenced architecturally.
We can leak this speculatively dereferenced value using the \RegAttack attack.

However, there are also many different causes for such gadgets~\cite{Intel2018retpoline}, \eg function pointers or (compiler-generated) jump tables.

\subsection{Generalization of Dereference Trap}
\RegAttack is a generic technique which also applies to any other scenario where the attacker can set up the hardware and address space accordingly.
\RegAttack applies to all Spectre variants. 
Thus, Spectre-v2 mitigations alone are not sufficient to hinder \RegAttack. 
Many in-place Spectre-v1 gadgets that are not the typical encoding array gadget are still entirely unprotected with no plans to change this.
For instance, Intel systems before Haswell and AMD systems before Zen do not support SMAP.
Also, more recent systems may have SMAP disabled.
On these systems, we can also \texttt{mmap} memory regions and the kernel will dereference 32-bit values misinterpreted as pointers (into user space).
We prepared an experiment where a kernel module speculatively accesses a secret value.
The user-space process performs the \RegAttack.
Using this technique the attacker can reliably leak a 32-bit secret which is speculatively dereferenced by the kernel module using an artificial Spectre gadget.
Cryptographic implementations often store keys in the lower 32 bits of 64bit registers (OpenSSL AES round key u32 *rk; for instance)~\cite{OpenSSL}.
Hence, those implementations might be susceptible to \RegAttack.

We evaluated the same experiment on an Intel i5-8250U, ARM Cortex-A57, and AMD ThreadRipper 1920X with the same result of $15$ minutes to recover a 32-bit secret.
Thus, Spectre-BTB mitigations and SMAP must remain enabled to mitigate attacks like \RegAttack.

\section{Leaking Physical Addresses from JavaScript using WebAssembly}\label{sec:js}
In this section, we present an attack that leaks the physical address (cache-line granularity) of a variable from within a JavaScript context.
Our main goal is to show that the ``prefetching'' effect is much simpler than described in the original paper~\cite{Gruss2016Prefetch}, \ie \emph{it does not require native code execution}.
The only requirement for the environment is that it can keep a 64-bit register filled with an attacker-controlled 64-bit value.

In contrast to the original paper's attempt to use NaCl to run in native code in the browser, we describe how to create a JavaScript-based attack to leak physical addresses from Javascript variables and evaluate its performance in common JavaScript engines and Firefox.
We demonstrate that it is possible to fill 64-bit registers with an attacker-controlled value in JavaScript by using WebAssembly.

\subparagraph{\textbf{Attack Setup.}}
JavaScript encodes numbers as double-precision floating-point values in the IEEE 754 format~\cite{JavascriptPrecision2019}.
Thus, it is not possible to store a full 64-bit value into a register with vanilla JavaScript, as the maximum precision is only 53-bit.
The same is true for Big-Integer libraries, which represent large numbers as structures on the heap~\cite{Google2019BigInt}.
To overcome this limitation, we leverage WebAssembly, a binary instruction format which is precompiled for the JavaScript engine and not further optimized by the engine~\cite{Google2019BigInt}.
The precompiled bytecode can be loaded and instantiated in JavaScript.
To prevent WebAssembly from harming the system, the bytecode is limited to calling functions provided by the JavaScript scope.

Our test operating system is Debian 8 (kernel\texttt{5.3.9-1kali1}) on an Intel i7-8550U.
We observe that on this system registers \texttt{r9} and \texttt{r10} are speculatively dereferenced in the kernel.
In our attack, we focus on filling these specific registers with a guessed direct-physical-map address of a variable.
The WebAssembly method \texttt{load\_pointer} of \Cref{lst:prefetchjs} (\Cref{sec:appendix-webassembly}) takes two 32-bit JavaScript values, which are combined into a 64-bit value and populated into as many registers as possible.
To trigger interrupts we rely on web requests from JavaScript, as suggested by Lipp~\etal\cite{Lipp2017Interrupt}.

We can use our attack to leak the direct-physical-map address of any variable in JavaScript.
The attack works analogously to the address-translation attack in native code~\cite{Gruss2016Prefetch}.
\begin{enumerate}[nolistsep,align=left, leftmargin=14pt, labelwidth=0pt, itemindent=!]
\item Guess a physical address $p$ for the variable and compute the corresponding direct-physical map address $d(p)$. 
\item Load $d(p)$ into the required registers (\texttt{load\_pointer}) in an endless loop, \eg using endless-loop slicing~\cite{Lipp2017Interrupt}.
\item The kernel fetches $d(p)$ into the cache when interrupted.
\item Use \EvictReload on the target variable.
On a cache hit, the physical address guess $p$ from Step 1 was correct.
Otherwise, continue with the next guess.
\end{enumerate}

\subparagraph{\textbf{Attack from within Browsers.}}
Before evaluating our attack in an unmodified Firefox browser, we evaluate our experiment on the JavaScript engines V8 version 7.7 and Spidermonkey 60.
To verify our experiments, we use Kali Linux (kernel \texttt{5.3.9-1kali1}) running on an Intel i7-8550U.
As it is the engines that execute our WebAssembly, the same register filling behavior as in the browser should occur when the engines are executed standalone.
In both engines, we use the C-APIs to add native code functions~\cite{Mozilla2019JSAPI,GoogleV8Blog2019}, enabling us to execute syscalls such as \texttt{sched\_yield}.
This shortcuts the search to find JavaScript code that constantly triggers syscalls.
Running inside the engine with the added syscall, we achieve a speed of $20$ speculative fetches per second. 
In addition to testing in the standalone JavaScript engines, we also show that speculative dereferencing can be triggered in the browser.
We mount an attack in an unmodified Firefox 76.0 by injecting interrupts via web requests.
We observe up to $2$ speculative fetches per hour.
If the logical core running the code is constantly interrupted, \eg due to disk I/O, we achieve up to $1$ speculative fetch per minute. 
As this attack leaks parts of the physical and virtual address, it can be used to implement various microarchitectural attacks~\cite{Oren2015,Pessl2016,Schwarz2017Timers,Gruss2016Row,Gras2017,Kocher2019,Schwarz2019STL}.
Hence, the address-translation attack is possible with JavaScript and WebAssembly, without requiring the NaCl sandbox as in the original paper~\cite{Gruss2016Prefetch}.

Upcoming JavaScript extensions expose syscalls to JavaScript~\cite{ChromeMojoAPI}.
However, at the time of writing, no such extensions are enabled by default.
Hence, as the second part of our evaluation, we investigate whether a syscall-based attack would also yield the same performance as in native code.
To simulate the extension, we expose the \texttt{sched\_yield} syscall to JavaScript.
We observe the same performance of $20$ speculative fetches per second with the syscall function.
Thus, new extensions for JavaScript may improve the performance of our previously described attack on unmodified Firefox.

\subparagraph{\textbf{Limitations of the Attack.}}
We conclude that the bottleneck of this attack is triggering syscalls.
In particular, there is currently no way to directly perform a single syscall via Javascript in browsers without high overhead.
We traced the syscalls of Firefox using \texttt{strace}.
We observed that syscalls such as \texttt{sched\_yield, getpid, stat, sendto} are commonly performed upon \texttt{window} events, \eg opening and closing pop-ups or reading and writing events on the JavaScript console.
However, the registers \texttt{r9} and \texttt{r10} get overwritten before the syscall is performed.
Thus, whether the registers are speculatively dereferenced while still containing the attacker-chosen values strongly depends on the engine's register allocation and on other syscalls performed.
As Jangda~\etal\cite{Jangda2019WebAssembly} stated, not all registers are used in Chrome and Firefox in the JIT-generated native code.
Not all registers can be filled from within the browser, \eg Chrome uses the registers \texttt{r10} and \texttt{r13} only as scratch registers, and Firefox uses \texttt{r15} as the heap pointer~\cite{Jangda2019WebAssembly}.


\section{Discussion}\label{sec:discussion}
The ``prefetching'' of user-space registers was first observed by Gruss~\etal\cite{Gruss2016Prefetch} in 2016.
In May 2017, Jann Horn discovered that speculative execution can be exploited to leak arbitrary data.
In January 2018, pre-prints of the Spectre~\cite{Kocher2019} and Meltdown~\cite{Lipp2018meltdown} papers were released.
Our results indicate that the address-translation attack was the first inadvertent exploitation of speculative execution, albeit in a much weaker form where only metadata, \ie information about KASLR, is leaked rather than real data as in a full Spectre attack.
Even before the address-translation attack, speculative execution was well known~\cite{Rebeiro2009} and documented~\cite{Intel_vol3} to cause cache hits on addresses that are not architecturally accessed.
This was often mentioned together with prefetching~\cite{Hund2013,Yarom2014Flush}.
Currently, the address-translation attack and our variants are mitigated on both Linux and Windows using the retpoline technique to avoid indirect branches.
In particular, the Spectre-BTB gadget in the syscall wrapper can be fixed by using the \texttt{lfence} instruction.

Another possibility upon a syscall is to save user-space register values to memory, clear the registers to prevent speculative dereferencing, and later restore the user-space values after execution of the syscall.
However, as has been observed in the interrupt handler, there might still be some speculative cache accesses on values from the stack.
The retpoline mitigation for Spectre-BTB introduces a large overhead for indirect branches.
The performance overhead can in some cases be up to \SI{50}{\percent}~\cite{Slashdot2019RetpolinePerformance}.
This is particularly problematic in large scale systems, \eg cloud data centers, that have to compensate for the performance loss and increased energy consumption.
Furthermore, retpoline breaks CET and CFI technologies and might thus also be disabled~\cite{Branco2019randpoline}.
As an alternative, randpoline~\cite{Branco2019randpoline} could be used to replace the mitigation with a probabilistic one, again with an effect on Foreshadow mitigations.
And indeed, mitigating memory corruption vulnerabilities may be more important than mitigating Foreshadow in certain use cases.
Cloud computing concepts that do not rely on traditional isolation boundaries are already being explored in industry~\cite{Amazon2019Lambda,Cloudflare2019Workers,Microsoft2019Azure,IBM2019CloudFunctions}.
Future work should investigate mitigations which take these new computing concepts into account rather than enforcing isolation boundaries that are less necessary in these use cases.

On current CPUs, Spectre-BTB mitigations, including retpoline, must remain enabled.
On newer kernels for ARM Cortex-A CPUs, the branch prediction results can be discarded, and on certain devices branch prediction can be entirely disabled~\cite{ARM2018Whitepaper}.
Our results suggest that these mechanisms are required for context switches or interrupt handling.
Additionally, the L1TF mitigations must be applied on affected CPUs to prevent Foreshadow.
Otherwise, we can still fetch arbitrary hypervisor addresses into the cache.
Finally, our attacks also show that SGX enclaves must be compiled with the retpoline flag.
Even with LVI mitigations, this is currently not the default setting, and thus all SGX enclaves which speculatively load secrets are potentially susceptible to \RegAttack.


\section{Conclusion}\label{sec:conclusion}
We confirmed the empirical results from several previous works~\cite{Gruss2016Prefetch,Lipp2018meltdown,Vanbulck2018foreshadow,Xiao2019Speechminer} while showing that the underlying root cause was misattributed in these works, resulting in incomplete mitigations~\cite{Gruss2017KASLR,Lipp2018meltdown,Vanbulck2018foreshadow,Canella2019A,Nilsson2020SGXSurvey,VanSchaik2019RIDL}.
Our experiments clearly show that speculative dereferencing of a user-space register in the kernel causes the leakage.
As a result, we were able to improve the performance of the original attack and show that CPUs from other hardware vendors like AMD, ARM, and IBM are also affected.
We demonstrated that this effect can also be exploited via JavaScript in browsers, enabling us to leak the physical addresses of JavaScript variables.
To systematically analyze the effect, we investigated its leakage capacity by implementing a cross-core covert channel which works without shared memory.
We presented a novel technique, \RegAttack, to leak the values of registers used in SGX (or privileged contexts) via speculative dereferencing.
We demonstrated that it is possible to fetch addresses from hypervisors into the cache from the guest operating system by triggering interrupts, enabling Foreshadow (L1TF) on data from the L3 cache.
Our results show that, for now, retpoline must remain enabled even on recent CPU generations to fully mitigate high impact microarchitectural attacks such as Foreshadow.

\section*{Acknowledgments}
We want to thank Moritz Lipp, Cl{\'e}mentine Maurice, Anders Fogh, Xiao Yuan, Jo Van Bulck, and Frank Piessens of the original papers for reviewing and providing feedback to drafts of this work and for discussing the technical root cause with us.
Furthermore, we want to thank Intel and ARM for valuable feedback on an early draft.
This project has received funding from the European Research Council (ERC) under the European Union’s Horizon 2020 research and innovation program (grant agreement No 681402). 
This work has been supported by the Austrian Research Promotion Agency (FFG) via the project ESPRESSO, which is funded by the province of Styria and the Business Promotion Agencies of Styria and Carinthia. 
Additional funding was provided by generous gifts from Cloudflare, from Intel, and from ARM. Any opinions, findings, and conclusions or recommendations expressed in this paper are those of the authors and do not necessarily reflect the views of the funding parties.

\bibliographystyle{ACM-Reference-Format}
\bibliography{main}

\FloatBarrier


\appendix
\section{Mistraining BTB for \texttt{sched\_yield}}
\label{sec:mistraining_yield}
We evaluate the mistraining of the BTB by calling different syscalls, fill all general-purpose registers with DPM address and call \texttt{sched\_yield}.
Our test system was equipped with Ubuntu 18.04 (kernel 4.4.143-generic) and an Intel i7-6700K.
We repeated the experiment by iterating over various syscalls with different parameters (valid parameters,NULL as parameters) $10$ times with \SIx{200000} repetitions.
\cref{tab:mistraining_btb} lists the best $15$ syscalls to mistrain the BTB when sched\_yield is performed afterwards.
On this kernel version it appears that the read and getcwd syscalls mistraing the BTB best if sched\_yield is called after the register filling.

\begin{table}[t]
  \caption{Table of syscalls which achieve the highest numbers of cache fetches, when calling sched\_yield after the register filling.}
\vspace{-0.1cm}
\setlength{\aboverulesep}{0pt}
\setlength{\belowrulesep}{0pt}
    \begin{center}
      \resizebox{\hsize}{!}{
      \begin{tabular}{ccc}
        \hline
        Syscall     & Parameters & Avg. \# cache fetches\\
        \hline
        readv       & readv(0,NULL,0);                                         & $13766.3$  \\
        getcwd      & syscall(79,NULL,0);                                      & $7344.7$   \\
        getcwd      & getcwd(NULL,0);                                          & $6646.9$   \\
        readv       & syscall(19,0,NULL,0);                                    & $5541.4$   \\
        mount       & syscall(165,s\_cbuf,s\_cbuf,s\_cbuf,s\_ulong,(void*)s\_cbuf); & $4831.6$   \\
        getpeername & syscall(52,0,NULL,NULL);                                 & $4600.0$   \\
        getcwd      & syscall(79,s\_cbuf,s\_ulong);                              & $4365.8$   \\
        bind        & syscall(49,0,NULL,0);                                    & $3680.6$   \\
        getcwd      & getcwd(s\_cbuf,s\_ulong);                                  & $3619.3$   \\
        getpeername & syscall(52,s\_fd,\&s\_ssockaddr,\&s\_int);                    & $3589.3$   \\
        connect     & syscall(42,s\_fd,\&s\_ssockaddr,s\_int);                     & $2951.2$   \\
        getpeername & getpeername(0,NULL,NULL);                                & $2822.4$   \\
        connect     & syscall(42,0,NULL,0);                                    & $2776.4$   \\
        getsockname & syscall(51,0,NULL,NULL);                                 & $2623.4$   \\
        connect     & connect(0,NULL,0);                                       & $2541.5$   \\
      \end{tabular}
    }
  \end{center}
  \label{tab:mistraining_btb}
\end{table}

\section{WebAssembly Register Filling}\label{sec:appendix-webassembly}

\begin{listing}[htb]
\begin{center}
\begin{lstlisting}[language=C]
extern void yield_wrapper();
uint64_t G1 = 5;
uint64_t G2 = 5;
uint64_t G3 = 5;
uint64_t G4 = 5;
uint64_t G5 = 5;
uint64_t value = 0;

void spec_fetch()
{
  for (uint64_t i = G1+5; i > G1; i--)
    for (uint64_t k = G3+5; k > G3; k--)
      for (uint64_t j = G2-5; k < G2; j++)
        for(uint64_t l = G4; i < G4;l++)
          for(uint64_t m = G5-5;m<G5;m++)
            value = l + j + k + i;
    yield_wrapper();
}

int load_pointer(int high, int low)
{
  uint64_t a = (((uint64_t)high) << 32ull) |
   ((uint64_t)(unsigned int)low);
  G1 = a;
  G2 = a;
  G3 = a;
  G4 = a;
  G5 = a;
  spec_fetch();
  return a;
}

int main()
{
  load_pointer(0x12345678,0x9abcdef0);
}
\end{lstlisting}
\vspace{-0.4cm}
\end{center}
\caption{WebAssembly code to speculatively fetch an address from the kernel direct-physical map into the cache.
We combine this with a state-of-the-art \EvictReload loop in JavaScript to determine whether the guess for the direct-physical map address was correct.}
\label{lst:prefetchjs}
\end{listing}

The WebAssembly method \texttt{load\_pointer} of \Cref{lst:prefetchjs} takes two 32-bit JavaScript values as input parameters.
These two parameters are loaded into a 64-bit integer variable and stored into multiple global variables.
The global variables are then used as loop exit conditions in the separate loops.
To fill as many registers as possible with the direct-physical-map address, we create data dependencies within the loop conditions.
In the \texttt{spec\_fetch} function, the registers are filled inside the loop.
After the loop, the JavaScript function \texttt{yield\_wrapper} is called.
This tries to trigger any syscall or interrupt in the browser by calling JavaScript functions which may incur syscalls or interrupts.
Lipp~\etal\cite{Lipp2017Interrupt} reported that web requests from JavaScript trigger interrupts from within the browser.
\section{No Foreshadow on Hyper-V HyperClear}
\label{sec:appendix}
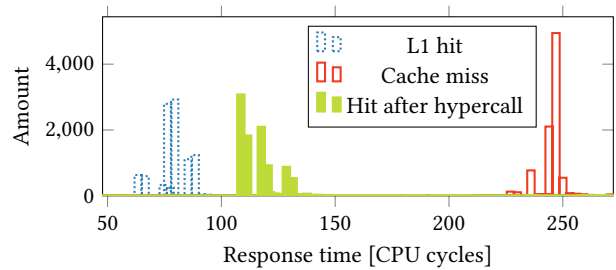
\begin{figure}[t]
    \centering
\resizebox{\hsize}{!}{
    \input{hist_hv.tikz}
    }
    \caption{Timings of a cached and uncached variable and the access time after a hypercall in a Ubuntu VM on Hyper-V.}
    \label{fig:access_time}
\end{figure}
We set up a Hyper-V virtual machine with a Ubuntu 18.04 guest (kernel \texttt{5.0.0-20}).
We access an address to load it into the cache and perform a hypercall before accessing the variable and measuring the access time.
Since hypercalls are performed from a privileged mode, we developed a kernel module for our Linux guest machine which performs our own malicious hypercalls.
We observe a timing difference (see~\Cref{fig:access_time}) between a memory access which hits in the L1 cache (dotted), a memory access after a hypercall (grid pattern), and an uncached memory access (crosshatch dots).
We observe that after each hypercall, the access times are approx. $20$ cycles slower.
This indicates that the guest addresses are flushed from the L1 data cache.
In addition, we create a second experiment where we load a virtual address from a process running on the host into several registers when performing a hypercall from the guest.
On the host system, we perform \FlushReload on the virtual address in a loop and verify whether the virtual address is fetched into the cache.
We do not observe any cache hits on the host process when performing hypercalls from the guest system.
Thus we conclude that either the L1 cache is always flushed, contradicting the documentation, or creating a situation where the L1 cache is not flushed requires a more elaborate attack setup.
However, we believe that speculative dereferencing is the reason why Microsoft adopted the retpoline mitigation despite having other Spectre-BTB mitigations already in place.
\end{document}

%% file: virt_to_phys.tikz
\begin{tikzpicture}[yscale=0.4,scale=0.85
]

\tikzstyle{box}+=[draw,thick]
\tikzstyle{asdf}+=[dotted,thick]

\draw[asdf] (0.75,5) -- (3,3.5) -- (3,0.5) -- (0.75,3.5);

\draw[box] (3.02,0.48) rectangle +(3.5,3) node[pos=.5] {Physical memory};

\draw[box,fill=black!5] (-3,0.5) rectangle +(3.75,1.5) node[pos=.5] {Non-canonical};

\draw[box] (-3,2) rectangle +(3.75,1.5) node[pos=.5] {Kernel};

\draw[box] (-3,3.5) rectangle +(3.75,1.5) node[pos=.5] {Direct-physical map};

\draw[box] (-3,-1) rectangle +(3.75,1.5) node[pos=.5] {User space};


\node at (-1.25,5.5) {Virtual memory per process};
\node at (-5.75,-1) {\texttt{0x0000\,0000\,0000\,0000}};
\node at (-5.75,0.5) {\texttt{0x0000\,8000\,0000\,0000}};
\node at (-5.75,3.5) {\texttt{0xffff\,8880\,0000\,0000}};
\node at (-5.75,2.) {\texttt{0xffff\,8000\,0000\,0000}};

\end{tikzpicture}

%% file: spec_acces_in_kernel.tikz
\begin{tikzpicture}[yscale=0.8]


\draw[fill=black!10] (3,3) rectangle +(3,1.5) node[pos=.5] {\parbox{2.5cm}{\centering Kernel}};
\draw[fill=red!10] (3,2.5) rectangle +(3,0.5) node[pos=.5] {\centering \small{indirect jmp}};

\draw[fill=blue!10] (0,5) rectangle +(3,0.5) node[pos=.5,yshift=0.cm] {\centering \small{VA}};

\draw (-4,4) rectangle +(3,0.5) node[pos=.5,yshift=0.cm] {\centering \small{DPM address}};
\draw (-4,2) rectangle +(3,0.5) node[pos=.5,yshift=-0cm] {\centering \small{DPM address}};
\draw (-4,2.5) rectangle +(3,0.5) node[pos=.5,yshift=-0cm] {\centering \small{DPM address}};
\draw (-4,3) rectangle +(3,0.5) node[pos=.5,yshift=-0cm] {\centering \small{DPM address}};
\draw (-4,3.5) rectangle +(3,0.5) node[pos=.5,yshift=-0cm] {\centering \small{DPM address}};

\draw (2.5,1) rectangle +(2.5,-2.5) node[pos=.5] {};
\draw (5.5,1) rectangle +(2.5,-2.5) node[pos=.5] {};
\draw (2.5,1) rectangle +(2.5,-.5) node[pos=.5] {\tiny{mov (\%rdx), \%rax}};

\draw (2.5,0.5) rectangle +(2.5,-.5) node[pos=.5] {\tiny{...}};
\draw (2.5,0.) rectangle +(2.5,-.5) node[pos=.5] {\tiny{...}};
\draw (2.5,-0.5) rectangle +(2.5,-.5) node[pos=.5] {\tiny{...}};
\draw (2.5,-1) rectangle +(2.5,-.5) node[pos=.5] {\tiny{...}};

\draw (5.5,1) rectangle +(2.5,-.5) node[pos=.5] {\tiny{...}};
\draw (5.5,0.5) rectangle +(2.5,-.5) node[pos=.5] {\tiny{...}};
\draw (5.5,0) rectangle +(2.5,-.5) node[pos=.5] {\tiny{...}};
\draw (5.5,-0.5) rectangle +(2.5,-.5) node[pos=.5] {\tiny{...}};
\draw (5.5,-1) rectangle +(2.5,-.5) node[pos=.5] {\tiny{...}};

\draw (0,5) edge[->,>=stealth,out=90,in=40,thick] (-2.5,4.5);

\draw (-1,2.7) edge[->,>=stealth,out=0,in=180,thick] (3,3);

\draw (2.5,0.5) edge[->,>=stealth,out=270,in=0,thick,looseness=0.7] (-2,0.3);

\draw[->] (4.5,2.5) -- node[above,rotate=60,xshift=-5] {\small{predicted}} (3.75,1);
\draw[->] (4.5,2.5) -- (6.75,1);

\draw (-4,-1) rectangle +(2,1.5) node[pos=.5,yshift=-.25cm] {\parbox{2cm}{}};
\draw[fill=yellow] (-4,.1) rectangle +(2,0.4) node[pos=.5] {cache line};

\node at (-4.5,4.25) {rax};
\node at (-4.5,3.75) {.};
\node at (-4.5,3.25) {.};
\node at (-4.5,2.75) {.};
\node at (-4.5,2.25) {r15};

\node at (-3.5,5.5) {\small{1. Fill registers with DPM address}};
\node at (1.0,3.25) {\small{2. Interrupt/Syscall}};
\node at (0.5,0.5) {\small{3. Cache fetch}};

\node at (3.75,-1.75) {\small{Handler A}};
\node at (6.75,-1.75) {\small{Handler B}};

\end{tikzpicture}

%% file: covert.tikz
\begin{tikzpicture}[yscale=0.6]

\draw[dotted] (-1,3.5) -- (0,3) -- (0,1.5) -- (-1,2) -- (-1,3.5);
\draw[fill=black!2,dotted] (3,3.5) -- (2,3) -- (2,1.5) -- (3,2) -- (3,3.5);

\draw (0,1.5) rectangle +(2,1.5) node[pos=.5,yshift=-.25cm] {\parbox{2cm}{\centering Page}};

\draw[fill=black!10] (3,2) rectangle +(2,1.5) node[pos=.5] {\parbox{2.5cm}{\centering User mapping\\\textit{v}}};

\draw (-3.5,2) rectangle +(2.5,1.5) node[pos=.5,yshift=-0.0cm] {\parbox{2.25cm}{\centering Direct-physical map \textit{p}}};

\draw (-1,3.3) edge[->,>=stealth,out=-20,in=170,thick] (0,2.82);
\draw (3,3.3) edge[->,>=stealth,out=200,in=10,thick] (2,2.82);

\draw[fill=yellow] (0,2.6) rectangle +(2,0.4) node[pos=.5] {cache line};

\node at (-2.5,1) {Sender};
\node at (2.5,1) {Receiver};

\draw[densely dotted,thick] (-0.5,0.75) rectangle (5.25,4);
\end{tikzpicture}

%% file: hv_l1tf.tikz
\begin{tikzpicture}[yscale=0.6]


\draw[fill=black!10] (0,0) rectangle +(3,1.5) node [pos=.5] {\parbox{2.5cm}{\centering Virtual machine}};

\draw (7,0) rectangle +(3,1.5) node[pos=.5,yshift=-0.1cm] {\parbox{2.5cm}{\centering Hypervisor}};

\draw[->,>=stealth,thick] (3,1) to node[midway,above] {Int./Hypercall with VA} (7,1) ;
\draw[->,>=stealth,out=270,in=195,thick,looseness=1.2] (1.5,0) to node[midway,below] {Foreshadow on PA} (4,0.25);
\draw[->,>=stealth,out=270,in=-15,thick,looseness=1.2] (9,0) to node[midway,below] {Fetch into cache} (6,0.25);

\draw (4,-1) rectangle +(2,1.5) node[pos=.5,yshift=-.25cm] {\parbox{2cm}{\centering Page}};
\draw[fill=yellow] (4,.1) rectangle +(2,0.4) node[pos=.5] {cache line};

\end{tikzpicture}

%% file: value_leak.tikz
\resizebox{\hsize}{!}{
\begin{tikzpicture}[yscale=0.55]
        \tikzset{fr/.style={anchor=south west,draw,minimum width=6.5cm,minimum height=0.6cm,align=center,draw=none,}}
        \tikzset{L1/.style={anchor=south west,draw,minimum width=4cm,minimum height=0.6cm,text width=4cm,align=center,fill=red!10}}
        \tikzset{L2/.style={anchor=south west,draw,minimum width=4cm,minimum height=0.6cm,text width=4cm,align=center,fill=blue!10}}
        \tikzset{L3/.style={anchor=south west,draw,minimum width=1.5cm,minimum height=0.6cm,text width=1cm,align=center,fill=red!10}}
        \tikzset{L4/.style={anchor=south west,draw,minimum width=1.5cm,minimum height=0.6cm,text width=1cm,align=center,fill=blue!10}}
        
        \node (fr0) at (-6,11) [fr] {\Large Flush+Reload};
        \node (shared1) at (-7,8.5) [L1] {\Large Physical Page $p1$};
        \node (shared2) at (-2.5,8.5) [L2] {\Large Physical Page $p2$};
        \begin{scope}[shift={(0,1)}]
        \node (s1mp1) at (-8.5,5.5) [L3] {\Large $v_0$};
        \node (s1mp2) at (-6.5,5.5) [L3] {\Large ...};
        \node (s1mp3) at (-4.5,5.5) [L3] {\Large $v_{\frac n 2 - 1}$};
        \node (s2mp1) at (-2.5,5.5) [L4] {\Large $v_{\frac n 2}$};
        \node (s2mp2) at (-0.5,5.5) [L4] {\Large ...};
        \node (s2mp3) at (1.5,5.5) [L4] {\Large $v_{n-1}$};
        
        \node [fr] (regval) at (-6,2.5) {\Large Register Value (between $v_0$ and $v_{n-1}$)};
        
        \draw[out=90,in=270,->,>=stealth,looseness=0.7] (s1mp1.north) to (shared1.south);
        \draw[out=90,in=270,->,>=stealth,looseness=0.8] (s1mp2.north) to (shared1.south);
        \draw[out=90,in=270,->,>=stealth,looseness=0.7] (s1mp3.north) to (shared1.south);

        \draw[out=90,in=270,->,>=stealth,looseness=0.7] (s2mp1.north) to (shared2.south);
        \draw[out=90,in=270,->,>=stealth,looseness=0.8] (s2mp2.north) to (shared2.south);
        \draw[out=90,in=270,->,>=stealth,looseness=0.7] (s2mp3.north) to (shared2.south);

        \draw[out=90,in=270,->,>=stealth,looseness=0.7,dotted] (regval.north) to (s1mp1.south);
        \draw[out=90,in=270,->,>=stealth,looseness=0.7,dotted] (regval.north) to (s1mp2.south);
        \draw[out=90,in=270,->,>=stealth,looseness=0.7,dotted] (regval.north) to (s1mp3.south);
        \draw[out=90,in=270,->,>=stealth,looseness=0.7,dotted] (regval.north) to (s2mp1.south);
        \draw[out=90,in=270,->,>=stealth,looseness=0.7,dotted] (regval.north) to (s2mp2.south);
        \draw[out=90,in=270,->,>=stealth,looseness=0.7,dotted] (regval.north) to (s2mp3.south);

        \draw[out=270,in=90,->,>=stealth,looseness=0.7,dashed] (fr0.south) to (shared1.north);
        \draw[out=270,in=90,->,>=stealth,looseness=0.7,dashed] (fr0.south) to (shared2.north);

        \node[color=black!60] at (-1.75,3.7) {Dereference};
        \end{scope}
        \node[color=black!60] at (-2,10.85) {Test};

\end{tikzpicture}
\vspace{-0.3cm}
}

%% file: hist_hv.tikz
\resizebox{\hsize}{!}{
\begin{tikzpicture}
            \begin{axis}[
            ybar,
            bar width=0.1cm,
            enlarge x limits={0.01},
            width={\hsize},
            height=4cm,
            xlabel={Response time [CPU cycles]},
            ylabel={Amount},
            xmin=50,
            xmax=270,
            ymin=0,
            legend style={at={(0.85,0.95)}}
            ]
            \addplot[thick,color=blue,densely dotted] table[x index = {0}, y index = {1}, col sep=comma]{hyperv-windows.csv};
            \addplot[thick,color=red]  table[x index = {0}, y index = {2}, col sep=comma]{hyperv-windows.csv};
            \addplot[thick,color=green,fill=green]  table[x index = {0}, y index = {3}, col sep=comma]{hyperv-windows.csv};
            \legend{L1 hit, Cache miss, Hit after hypercall}
            \end{axis}
\end{tikzpicture}
}